\documentclass{aa}
\usepackage{natbib}
\usepackage{graphicx}
\usepackage{lipsum}
\usepackage{multirow}

\usepackage{commath}
\usepackage[switch,columnwise]{lineno} 

\usepackage[super]{nth}
\usepackage{tikz}
\usetikzlibrary{shapes,arrows}
\usepackage{array}
\usepackage{subcaption}
\usepackage[varg]{txfonts}
\usepackage{amsmath}
\usepackage{booktabs}
\usepackage{placeins}

\usepackage{longtable}
\usepackage{geometry}
\usepackage{caption}
\usepackage{adjustbox}
\usepackage{float}
\usepackage{tabularx} 

\usepackage[varg]{txfonts}
\usepackage{commath}
\usepackage[super]{nth}

\usepackage{siunitx}
\definecolor{click_color}{RGB}{46,48,118}

\usepackage{mathrsfs}

\usepackage{graphicx}
\graphicspath{{figures/}}
\usepackage[varg]{txfonts}
\usepackage{commath}
\usepackage[super]{nth}

\usepackage{siunitx}
\DeclareSIUnit{\erg}{erg}
\DeclareSIUnit{\jansky}{Jy}
\DeclareSIUnit{\parsec}{pc}

\usepackage{hyperref}
\hypersetup{
  colorlinks=true,
  linkcolor=click_color,
  citecolor=click_color,
  urlcolor=click_color
  }
\usepackage[capitalize]{cleveref}
\crefname{equation}{eq.}{eqs.}
\crefname{section}{Sect.}{Sects.}

\usepackage{txfonts}

\usepackage[normalem]{ulem}

\begin{document} 

\newcommand*{\frbA}[1]{FRB20240114A~#1}
\newcommand*{\frbB}[1]{FRB20121102A~#1}
\newcommand*{\frbC}[1]{FRB20180916B~#1}
\newcommand*{\frbD}[1]{FRB20201124A~#1}
\newcommand{\scintBW}{\ensuremath{\Delta \nu_{\rm scint}}}
\newcommand{\emitBW}{\ensuremath{\Delta \nu_{\rm emit}}}

   \title{A broadband study of \frbA with the Effelsberg 100-m radio telescope }

\author{
    P. Limaye\thanks{Email: limaye@mpifr-bonn.mpg.de}\inst{1,2}
    \and
    L.G. Spitler\inst{1}
    \and
    N. Manaswini\inst{1}
    \and
    J. Benáček\inst{4}
    \and
    F.\,Eppel \inst{3,1,5}
    \and
    M.\,Kadler \inst{3}
    \and
    L. Nicotera\inst{1}
    \and
    J.\,Wongphechauxsorn \inst{3,1}
    }

\institute{
     Max-Planck-Institut für Radioastronomie, Auf dem Hügel 69, 53121 Bonn, Germany
    \and
     Argelander Institute for Astronomy, Auf dem Hügel 71, 53121 Bonn, Germany
    \and
    Julius-Maximilians-Universität Würzburg, Institut für Theoretische Physik und Astrophysik, Lehrstuhl für Astronomie, Emil-Fischer-Straße 31, 97074 Würzburg, Germany
    \and
    Institute for Physics and Astronomy, University of Potsdam, 14476 Potsdam, Germany
    \and
    Joint Institute for VLBI ERIC, Oude Hoogeveensedĳk 4, 7991 PD Dwingeloo, The Netherlands
    }

\date{Received XXXX; accepted XXXX}


\abstract
{\frbA is a hyperactive repeating FRB discovered by the CHIME/FRB Collaboration in January 2024. The source has been followed up with numerous radio telescopes, including MeerKAT, uGMRT, and FAST. The FRB has been localized to a dwarf star-forming galaxy at a redshift of $z \sim 0.13$ and has a confirmed persistent radio source.
}
{In this study, we report the observations of \frbA with the Effelsberg 100-m radio telescope. Observations were conducted using the Ultra BroadBand (UBB) receiver, which offers a unique frequency coverage from 1.3 to 6.0~GHz, allowing us to detect over 700 unique bursts from this source. The observations were conducted over four epochs, spaced roughly two weeks apart, except for the first epoch, which was scheduled within a month of the source’s discovery.
}
{Using the observed sample of bursts, we performed an extensive statistical study of the broadband-burst properties. This includes an investigation of their morphologies, occurrence rates, spectral and temporal widths, and waiting-time distributions as a function of frequency across six sub-bands spanning the UBB frequency range.
}
{We classified the burst spectra into four distinct morphologies based on their spectro-temporal structures, which exhibit simple, complex, or frequency-drifting patterns. In the broadband observations, we detected no bursts across the full 1.3-6~GHz band, confirming the band-limited emission seen in other repeating FRBs. We observe modest frequency-dependent evolution in burst widths but no frequency evolution in their fractional bandwidths. The burst rates very significantly in time and frequency over the four epochs, but note that scintillation could play a significant role in the observed variability. The waiting-time distributions suggest that the bursts are largely independent events, though we also detect instances of temporal clustering that may point to a characteristic emission timescale ($\sim$10~ms).
Additionally, this study presents a multi-frequency analysis of waiting-time distributions, offering new insights into the complex frequency drifts commonly observed in repeating FRBs.
}
{}


%
\maketitle

\section{Introduction}

Repeating Fast Radio Bursts (FRBs) have emerged as a distinct subclass within the broader FRB population, with an increasing number of sources discovered in recent years through surveys conducted using telescopes such as CHIME, MeerKAT, the Australian Square Kilometre Array Pathfinder (ASKAP), and the Deep Synoptic Array (DSA-110) \citep{chime_frb_survey, meerkat_frb_survey, ASKAP_Discoveries, DSA_Discoveries}. The first known repeating FRB, \frbB, was discovered by \citet{Spitler_oneoff} as part of the PALFA survey with the Arecibo radio telescope. This source was later identified as part of a subclass of highly active repeaters, producing thousands of bursts during episodes of high activity \citep[see][]{jahns, Hewitt_R1_burststorm}. Of the $\sim900$ FRBs discovered to date\footnote{\url{https://ecommons.cornell.edu/server/api/core/bitstreams/a5fa037c-8edc-4e47-8154-df2fb70344ee/content}}, repeating bursts have been observed from $\sim$57 sources, and just a handful of these exhibit frequent, active repetition. Expanding the known sample of repeating FRBs is therefore critical for developing a comprehensive understanding of FRB physical properties and origins.

Individual bursts from repeating FRBs tend to exhibit narrower emission bandwidths than those from non-repeating, or ``one-off,'' FRBs \citep{ziggy}. Despite their typically narrow emission bandwidths, repeaters have been detected across a broad frequency range, from as low as 110~MHz to as high as 8.0~GHz \citep{ziggy_110MHz, 8ghz_frb, R3_Bethapudi}. Yet, even nearly a decade after the discovery of the first repeater, the nature of their progenitors remains elusive. Leading models invoke highly magnetized neutron stars, particularly magnetars, as the central engine \citep{Kumar2017, LuKumar2018}, though it remains unclear whether the emission could arise from coherent processes within the magnetosphere or from external shock interactions in the surrounding plasma \citep{Metzger2019, Beloborodov2020, Margalit2020}. As upcoming surveys promise to grow the sample of known repeaters, assembling statistically significant datasets will be essential for disentangling their emission physics. In particular, broadband and sensitive observations are key to probe their spectral and temporal properties in greater detail.

Previous studies have shown that detailed investigations of burst morphologies, rates, waiting times, and energy distributions can offer valuable insights into the emission mechanisms of repeating FRBs (see for eg. \citealp{R3_Bethapudi, jahns, cruces2021}). Notably, complex frequency-time structures, such as downward-drifting bursts, appear as recurring features among multiple repeaters \citep{ziggy}. Analyses of waiting times reveal short-timescale clustering, which suggests the presence of a characteristic emission timescale. At longer timescales, the behavior is well described by an exponential distribution, consistent with a Poissonian process. Potential periodicities or quasi-periodicities may be embedded within this process \citep{jahns, cruces2021}. Although some sources, such as \frbC and \frbB, have been studied at multiple radio frequencies \citep{R3_Bethapudi, 8ghz_frb}, these investigations are often constrained by the limited bandwidths of individual receivers and the lack of overlapping frequency coverage. In particular, \citet{R3_Bethapudi} demonstrated that \frbC exhibits strong frequency-dependent burst activity, emphasizing the need to capture its full spectral behavior. These limitations underscore the importance of coordinated, simultaneous multi-frequency observations of active repeaters.

To further investigate these questions, we conducted follow-up observations of the active repeating \frbA using the 100-m Effelsberg Radio Telescope. This hyperactive repeater was initially discovered by the CHIME/FRB Collaboration at a central frequency of 600~MHz, with a reported dispersion measure (DM) of $527.7~\mathrm{pc~cm^{-3}}$ \citep{chime_atel}. Following its discovery, multiple radio facilities reported detections at or below 1.4~GHz \citep{panda_r147_paper, R147_WaitTime_FAST, atel_parkes}, while continued monitoring revealed increased activity at frequencies above 2~GHz \citep{atel_nancay, atel_allentelescope, atel}. The source was first localized to the host galaxy J212739.84+041945.8 by MeerKAT \citep{meerkat_paper}, and later refined by the European VLBI Network (EVN) PRECISE team, which pinpointed the position to RA = 21:27:39.835, Dec = +04:19:45.634 \citep{atel_localisation_evn}, consistent with the MeerKAT coordinates. Optical follow-up observations confirmed the redshift of the host galaxy to be $z = 0.13$ \citep{atel_redshift}. Subsequent observations with the Very Long Baseline Array (VLBA) established an association between the FRB and a persistent continuum radio source (PRS), making it the fourth known FRB linked to such a source \citep{R147_PRS}.

The paper is structured as follows: In Section~\ref{sec:obs_search}, we describe the observational setup, single-pulse search strategy, calibration procedures, and data analysis methods used to extract the properties of individual bursts. In Section~\ref{sec:results}, we present the results of our study and discuss their implications. Finally, Section~\ref{sec:conclusion} summarizes our findings and offers interpretations based on the observational outcomes.

\section{Observations and data recording}
\label{sec:obs_search}

The CHIME/FRB collaboration discovered \frbA after detecting multiple bright bursts within a few days, prompting follow-up observations with the Effelsberg radio telescope. 

We monitored the source across four observing epochs, with the third epoch conducted simultaneously with \textit{XMM-Newton}, enabling constraints on its X-ray to radio fluence ratio \citep{eppel_r147}. Observations targeted the MeerKAT position of the source (RA = 21:27:39.83, Dec = +04:19:46.02, \citealp{meerkat_paper}). 

Data were recorded using the UBB receiver, covering 1.3--6.0~GHz with a 2.6--3.0~GHz frequency gap due to instrumental limitations. The EDD backend \citep{EDD} produced 8-bit full-Stokes \texttt{PSRFITS} search-mode files across five sub-bands within the usable range. The first two epochs used a native time resolution of 64~$\mu$s, and the last two 128~$\mu$s. All data were coherently dedispersed at the MeerKAT-derived dispersion measure of 527.7~pc~cm$^{-3}$ \citep{meerkat_paper}. Further details of the observing setup are provided in Appendix Table~\ref{table:rec_details}.

Flux calibration was performed using observations of the standard calibrator source 3C48 with the \texttt{PSRCHIVE} software suite. Calibrator observations were conducted only during one epoch (MJD~60439), but the Effelsberg system exhibits stable systematics, so the same calibration solution was applied to all epochs. The resulting system equivalent flux densities (SEFDs) are summarized in Appendix Table~\ref{table:rec_details}.

\subsection{Single pulse search}

Each EDD sub-band \texttt{PSRFITS} file was searched independently using the single-pulse search software \texttt{TransientX} \citep{TransientX}, which is optimized for efficient processing of large datasets. A detection threshold of $\mathrm{S/N} > 7$ was applied, and the resulting candidates were refined using \texttt{replot\_fil} before final visual inspection.  The full search procedure, including details of downsampling, RFI mitigation, dedispersion, and clustering strategies, is described in Appendix~\ref{app:SPsearch}.  

Since each EDD sub-band was searched independently, a single burst could appear in multiple adjacent sub-bands. To identify unique events, all detections were referenced to the top of the UBB band (6.0~GHz). Assuming a temporal uncertainty of approximately 10~ms—comparable to the typical burst duration—we clustered detections across sub-bands occurring within this window. This procedure yielded a final list of unique bursts for each observing epoch, as summarized in Table~\ref{table:detections}. We note that the burst detection statistics reported in this work differ slightly from those presented by \citet{eppel_r147}, as we adopt a more conservative selection to enable robust statistical analysis (see Section~\ref{sec:burst_rate}).

\begin{table*}[h!]
\centering
\begin{tabular}{lccccc}
\toprule
\textbf{Epoch (MJD)} & \textbf{Start Time (UTC)} & \textbf{Label} & \textbf{Exposure} & \textbf{Sampling Rate} & \textbf{Detections} \\
 &  &  & (hrs) & ($\mu$s) & (No.) \\
\midrule
60356.5109134 & 2024-02-16 12:15:42.918 & Epoch 1 & 1.8 & 64  & 5 \\
60439.1703909 & 2024-05-09 04:05:21.774 & Epoch 2 & 4.7 & 64  & 249 \\
60452.9943398 & 2024-05-22 23:51:50.959 & Epoch 3 & 6.2 & 128 & 436 \\
60471.1849648 & 2024-06-10 04:26:20.959 & Epoch 4 & 2.8 & 128 & 104 \\
\bottomrule
\end{tabular}

\caption{Effelsberg observations of \frbA conducted using UBB across the four epochs. The table lists the central Modified Julian Date (MJD), start time, on-source exposure, sampling rate, and the number of detected bursts.}
\label{table:detections}
\end{table*}

\subsection{Burst properties}
\label{sec:burst_properties}

We extracted burst Time-Of-Arrivals (TOAs) from \texttt{TransientX} candidates, dedispersed to a fixed DM of 527.979~pc~cm$^{-3}$ using \texttt{DSPSR}. To mitigate RFI, we applied manual masks, a zero-DM filter (up to Band~3), and \texttt{CLFD} \citep{CLFD}. Flux calibration was performed using observations of the quasar 3C48, processed via \texttt{PSRCHIVE}.

\begin{figure*}[h!]
  \centering

  \begin{subfigure}[t]{0.48\textwidth}
    \centering
    \includegraphics[width=0.49\textwidth]{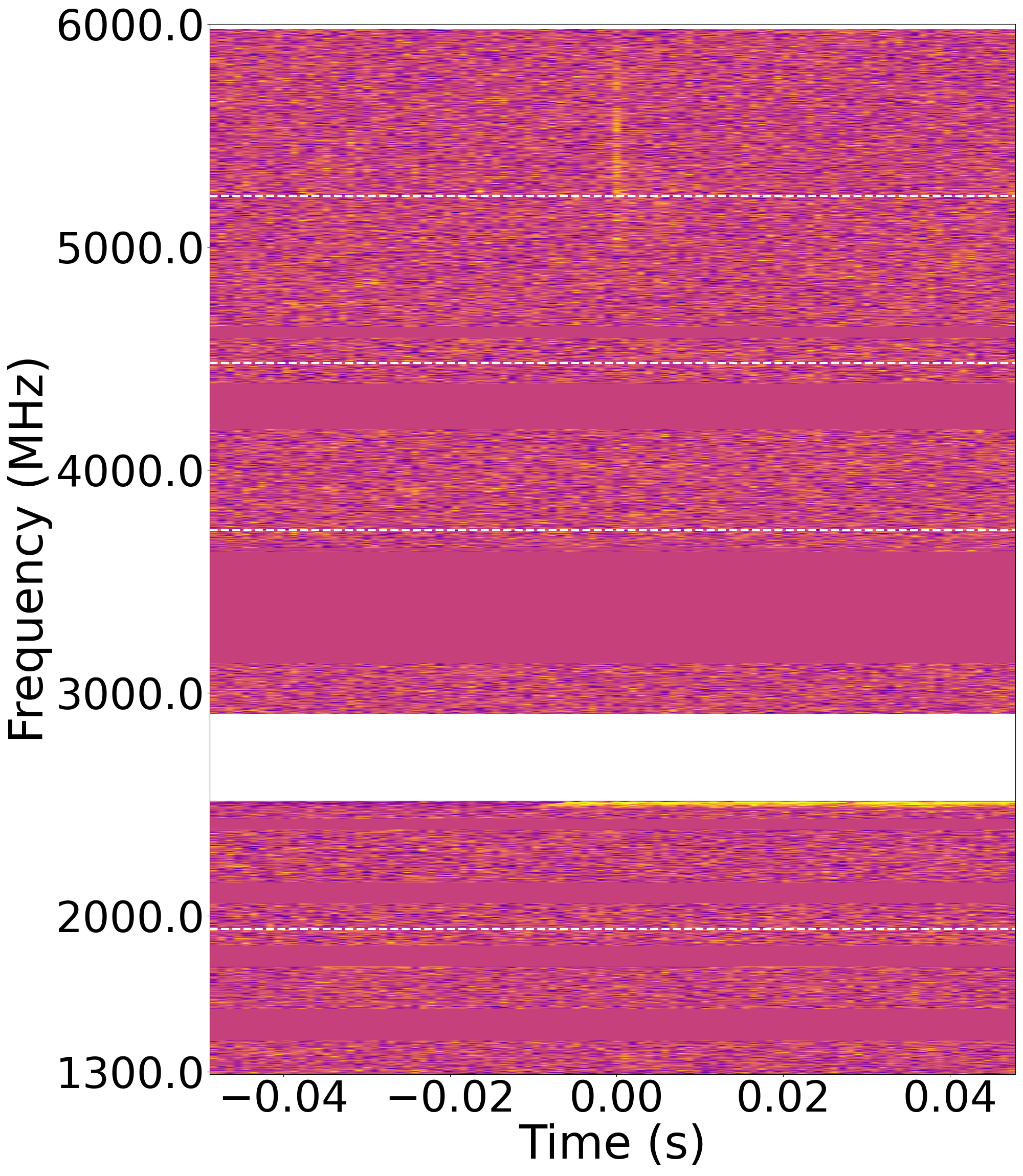}
    \includegraphics[width=0.49\textwidth]{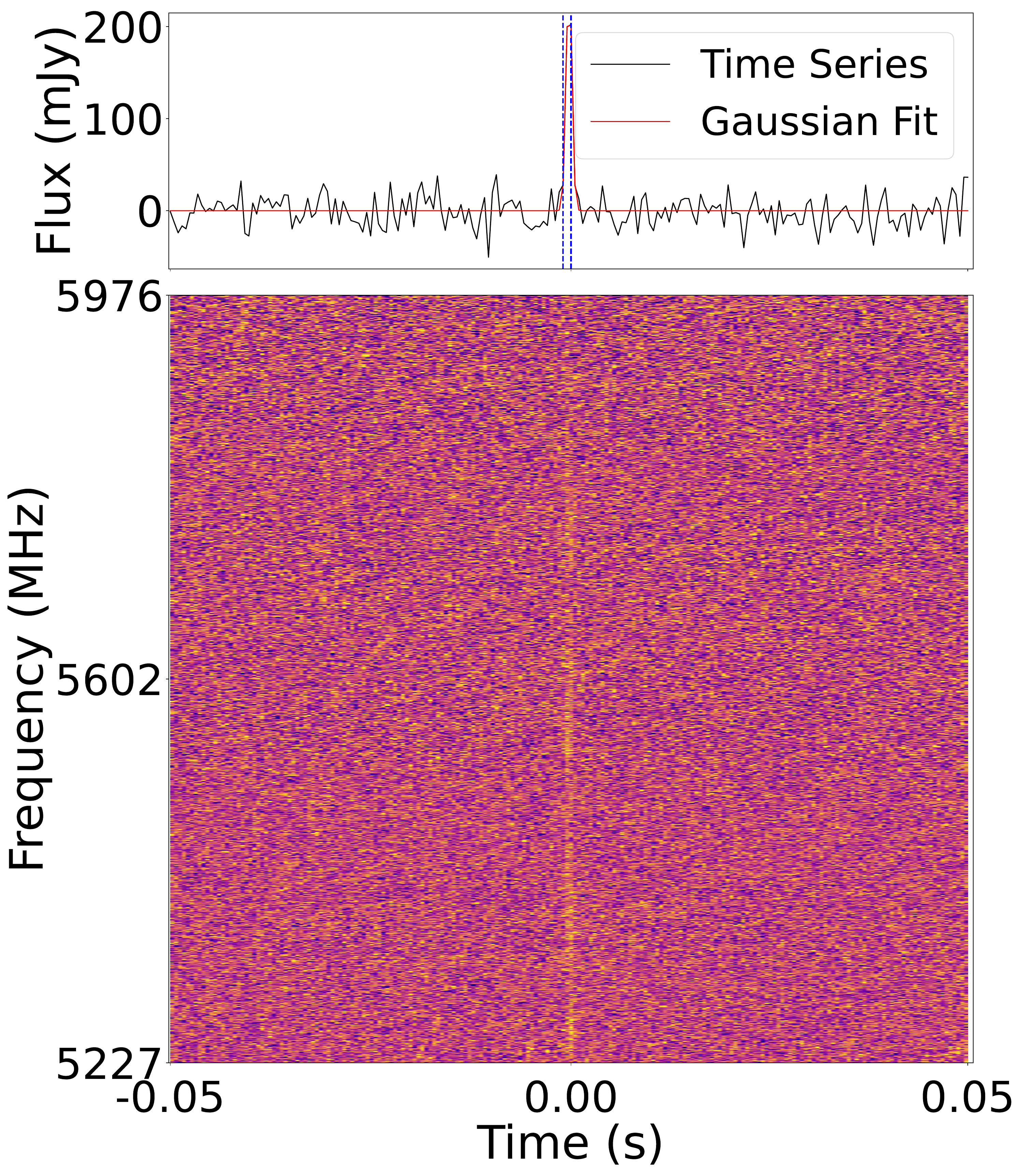}
    \caption{Simple narrowband}
  \end{subfigure}%
  \hspace{1em}
  \begin{subfigure}[t]{0.48\textwidth}
    \centering
    \includegraphics[width=0.49\textwidth]{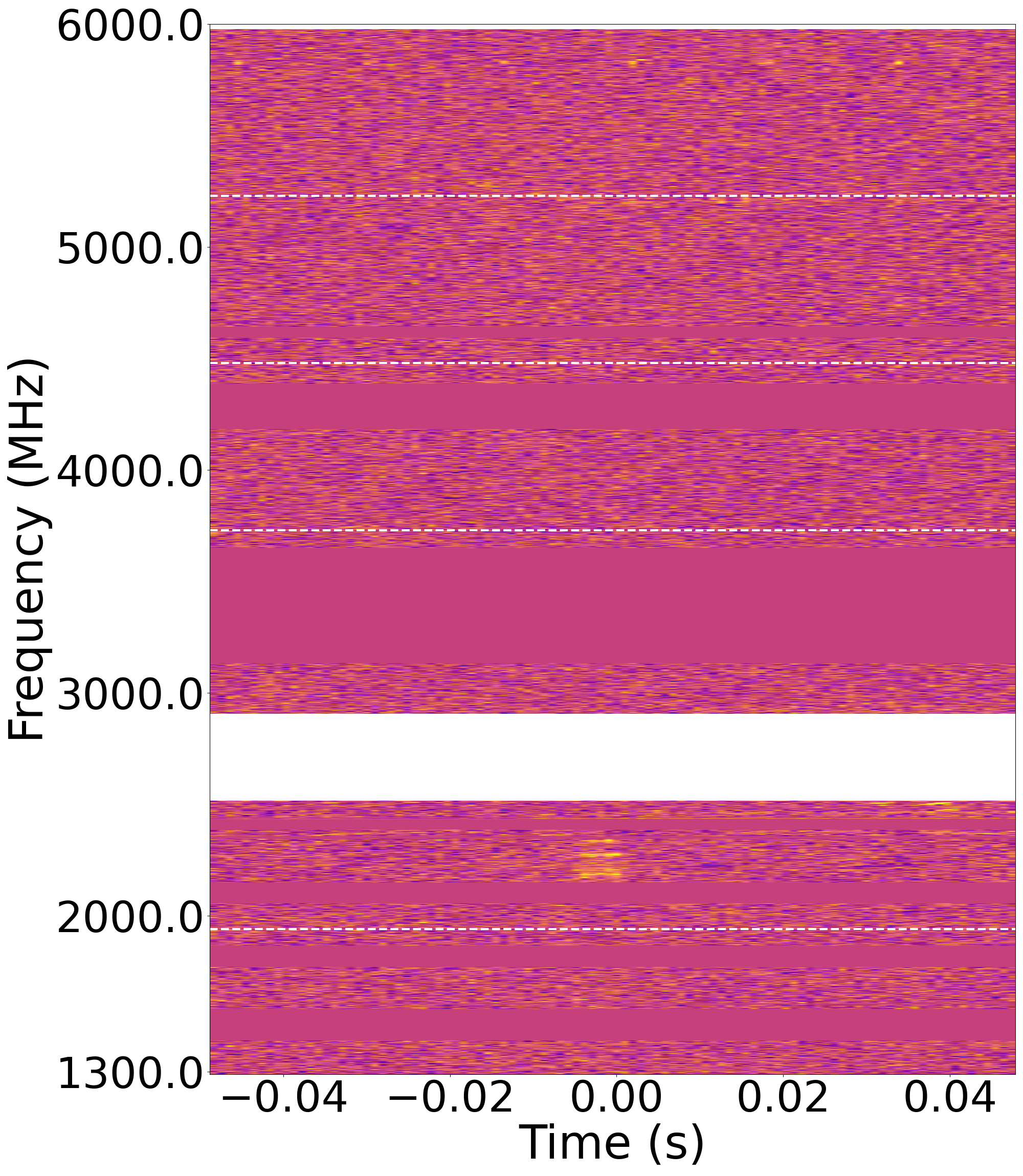}
    \includegraphics[width=0.49\textwidth]{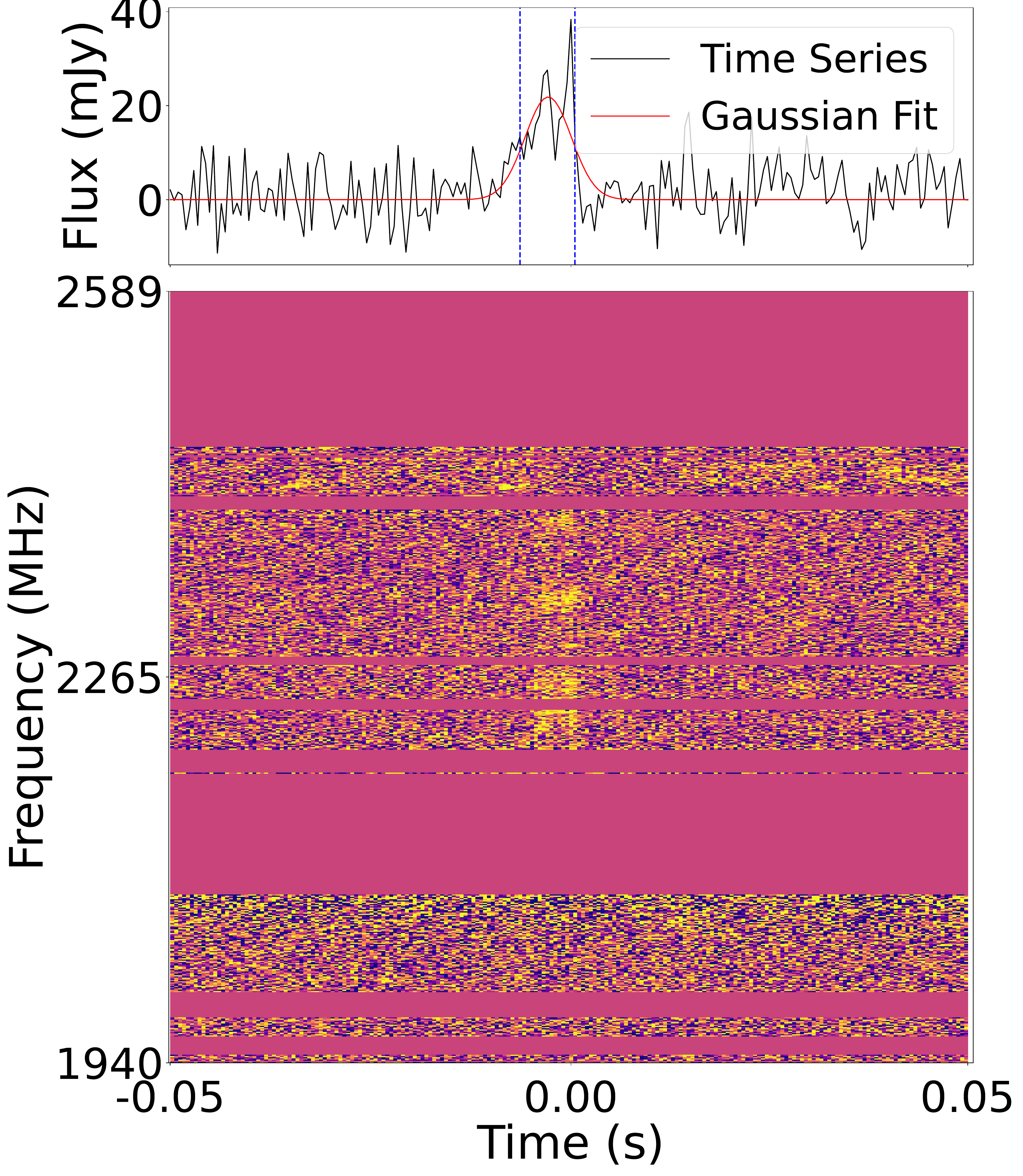}
    \caption{Complex multi-component}
  \end{subfigure}%

  \vspace{1em}

  \begin{subfigure}[t]{0.48\textwidth}
    \centering
    \includegraphics[width=0.49\textwidth]{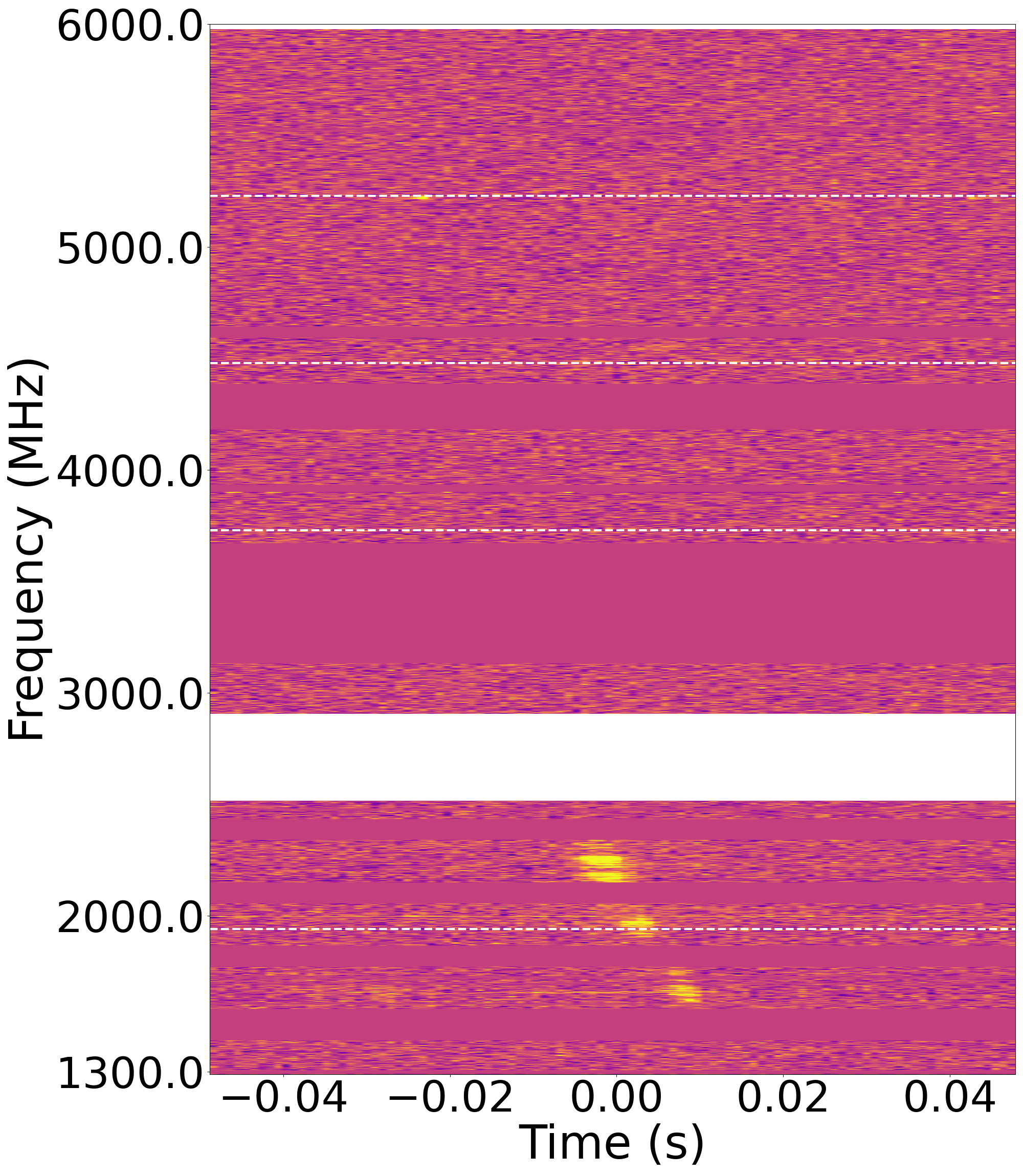}
    \includegraphics[width=0.49\textwidth]{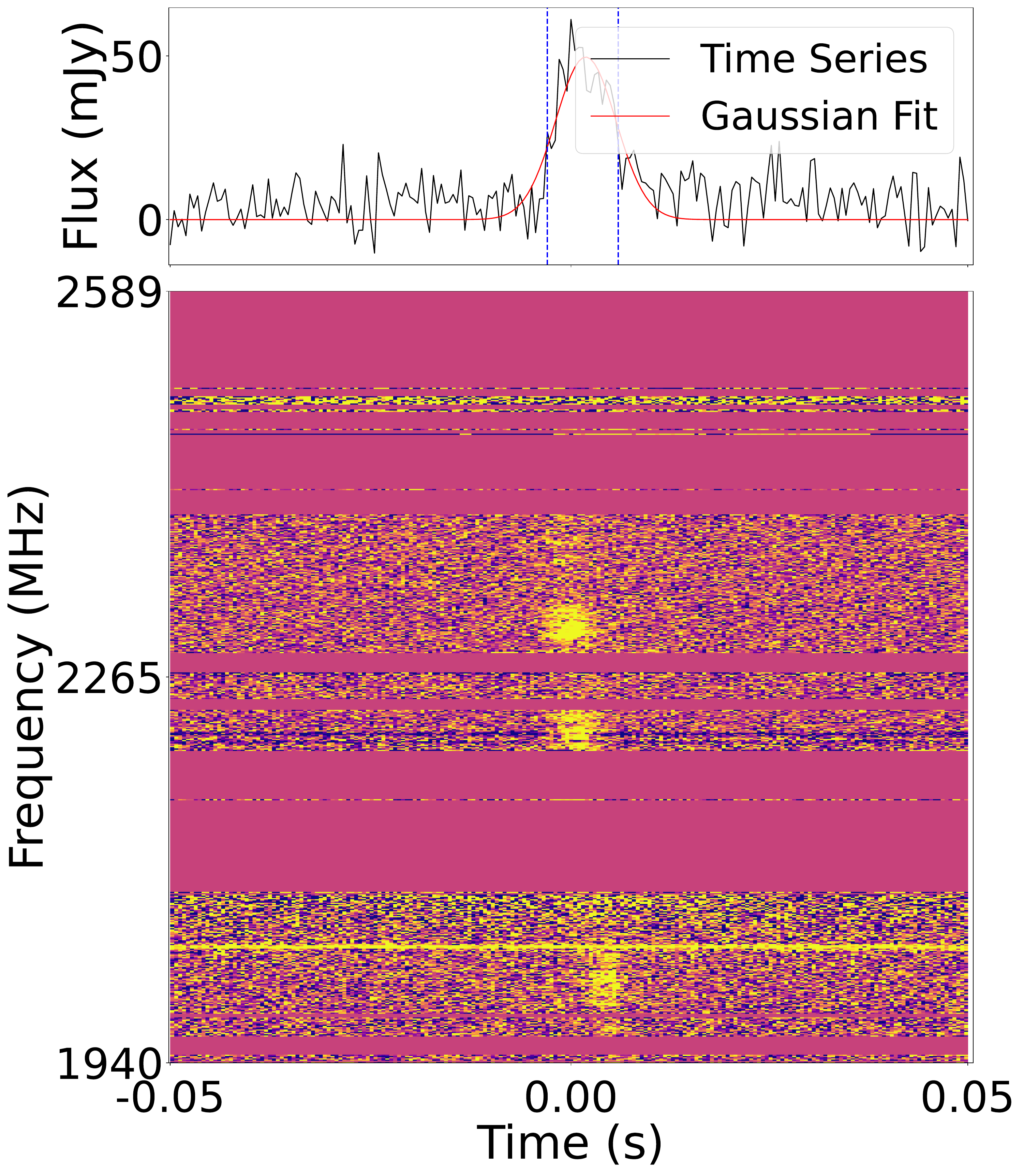}
    \caption{Downward frequency drifting}
  \end{subfigure}%
  \hspace{1em}
  \begin{subfigure}[t]{0.48\textwidth}
    \centering
    \includegraphics[width=0.49\textwidth]{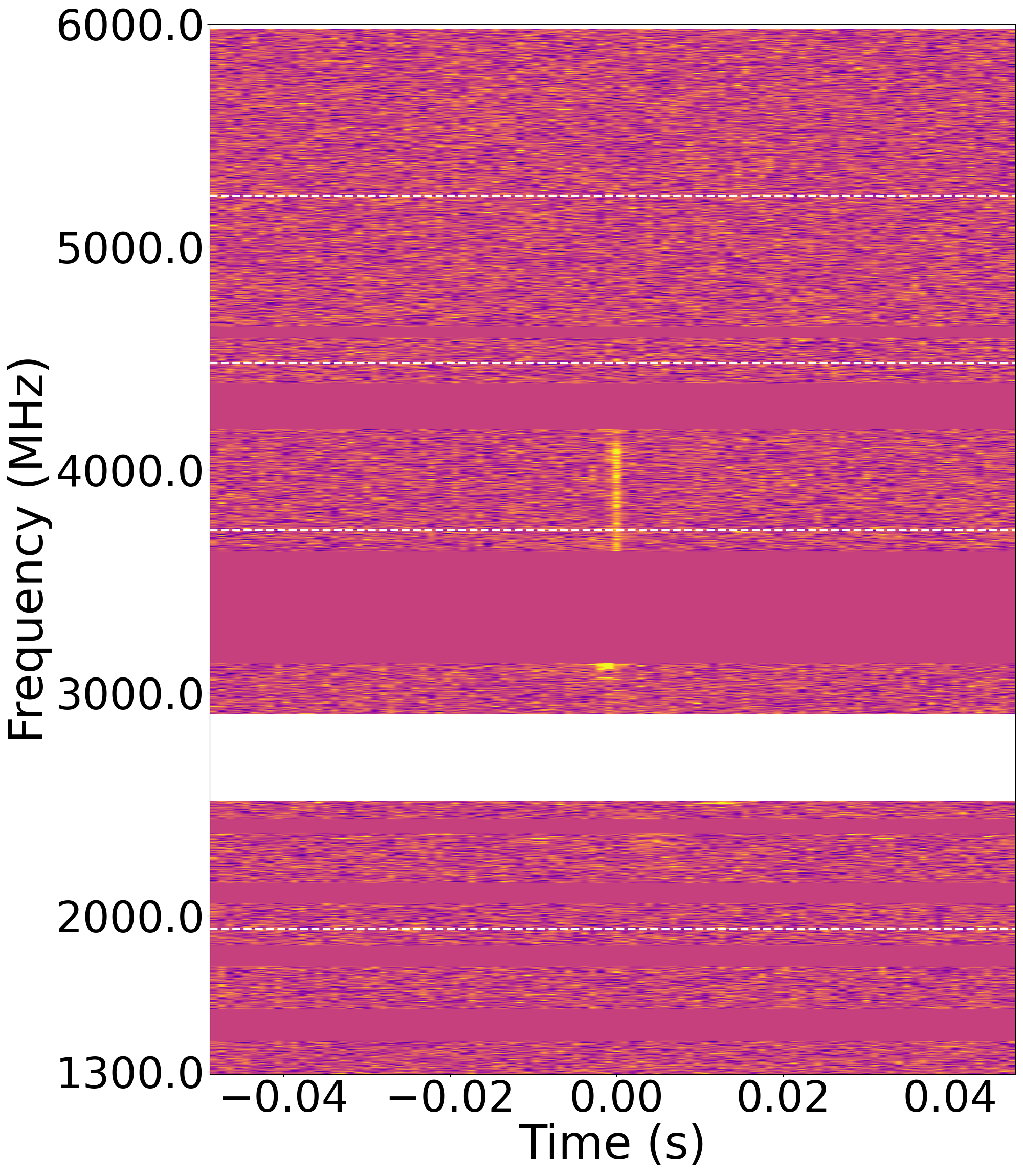}
    \includegraphics[width=0.49\textwidth]{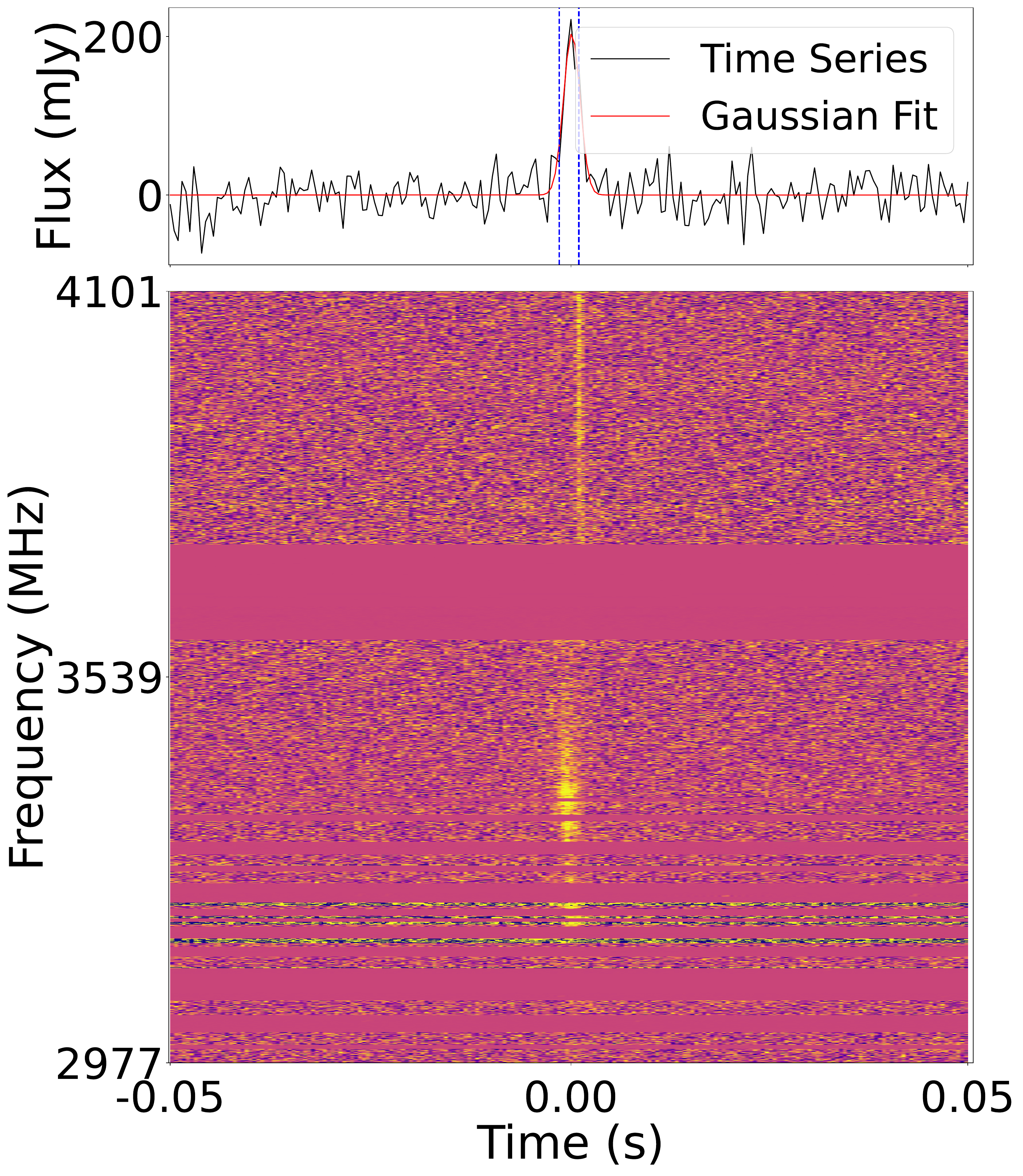}
    \caption{Upward frequency drifting}
  \end{subfigure}%

  \caption{Example dynamic spectra of selected bursts from \frbA, illustrating the distinct morphologies discussed in Section~\ref{sec:spec-temp_properties}. In each subplot, the left panel shows the burst spectrum across the full UBB frequency coverage where the vertical dashed lines indicate the rebinned frequency bands, while the right panel presents the dynamic spectrum (bottom) in the UBB band where the burst is detected, together with its frequency-averaged temporal profile (top). The temporal profiles are fitted with red solid curves, with dashed vertical lines marking the measured burst widths. The burst spectra were generated using a frequency downsampling factor of 2 and a time resolution of 0.5~ms. Channels affected by RFI were masked and replaced by median values for visualization. See Section~\ref{sec:burst_properties} for details of the fitting methodology.}
  \label{fig:burst_spectra}
\end{figure*}

Burst dynamic spectra and temporal profiles were reconstructed from calibrated archives (see Fig.~\ref{fig:burst_spectra}). Bandwidths were visually estimated due to RFI contamination in lower bands. A multi-Gaussian fitting procedure was applied to characterize burst widths, sub-components, TOAs, and fluences. Further methodological details and a catalog of derived properties are provided in Appendix~\ref{app:burst_appendix}.

\section{Results and discussion}
\label{sec:results}

As described in Table~\ref{table:rec_details}, the UBB sub-bands have different bandwidths. To enable a frequency-dependent analysis, we re-binned the dynamic spectra into six bands (RB1 to RB6) with approximately comparable bandwidths (see Table~\ref{table:rebinned_bands}).

This scheme provided an integer number of bands while ensuring a sufficient number of bursts in each band for meaningful statistics. The SEFDs at these binned bands were computed by interpolating the SEFDs derived for the native bands from the flux calibration. The bursts were assigned to a rebinned band if their center frequency lay within the frequency range of that band. Owing to the instrumental band gap mentioned in Section~\ref{sec:obs_search}, bursts with center frequencies falling within this gap were assigned to either RB2 or RB3, depending on which band was closer to the burst's estimated center frequency.

\begin{table*}[t]
\centering
\begin{tabular}{lccccc}
\toprule
\textbf{Re-binned Band} & \textbf{Frequency Range} & \textbf{Bandwidth} & \textbf{Center Frequency} & \textbf{SEFD} & \textbf{Fluence Completeness} \\
 & (MHz) & (MHz) & (MHz) & (Jy) & (Jy\,ms) \\
\midrule
RB1 & 1290--1940 & 650 & 1615 & 12.2 & 0.11 \\
RB2 & 1940--2590 & 650 & 2265 & 11.2 & 0.14 \\
RB3 & 2980--3730 & 750 & 3355 & 11.5 & 0.09 \\
RB4 & 3730--4480 & 750 & 4105 & 12.7 & 0.11 \\
RB5 & 4480--5230 & 750 & 4855 & 14.6 & 0.11 \\
RB6 & 5230--5980 & 750 & 5605 & 17.5 & 0.14 \\
\bottomrule
\end{tabular}
\caption{Frequency sub-bands and interpolated SEFD values used in the spectral analysis of bursts detected with the UBB receiver. These re-binned intervals were defined to ensure approximately uniform spectral sensitivity and consistent bandwidths for comparative studies.}
\label{table:rebinned_bands}
\end{table*}

\subsection{Spectro-Temporal properties}
\label{sec:spec-temp_properties}

\subsubsection{Burst morphologies}
We manually classified the bursts into four primary morphological categories based on their frequency--time structure, as illustrated in Fig.~\ref{fig:burst_spectra}: (1) simple narrowband, (2) complex multi-component, (3) downward frequency drifting, and (4) upward frequency drifting.

\begin{figure*}
    \centering
    \includegraphics[width=0.29\linewidth]{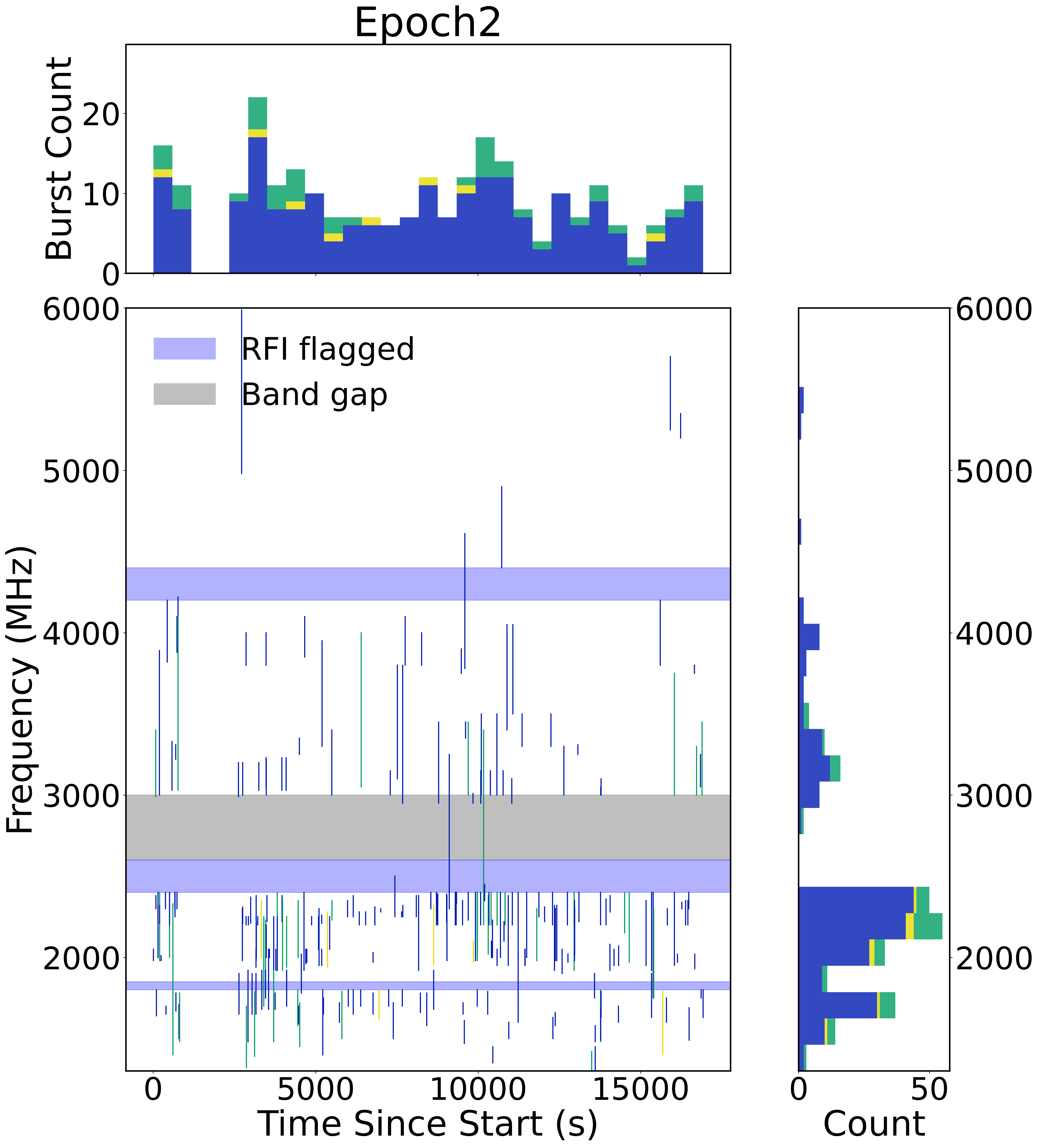}
    \hspace{0.01\linewidth}%
    \includegraphics[width=0.32\linewidth]{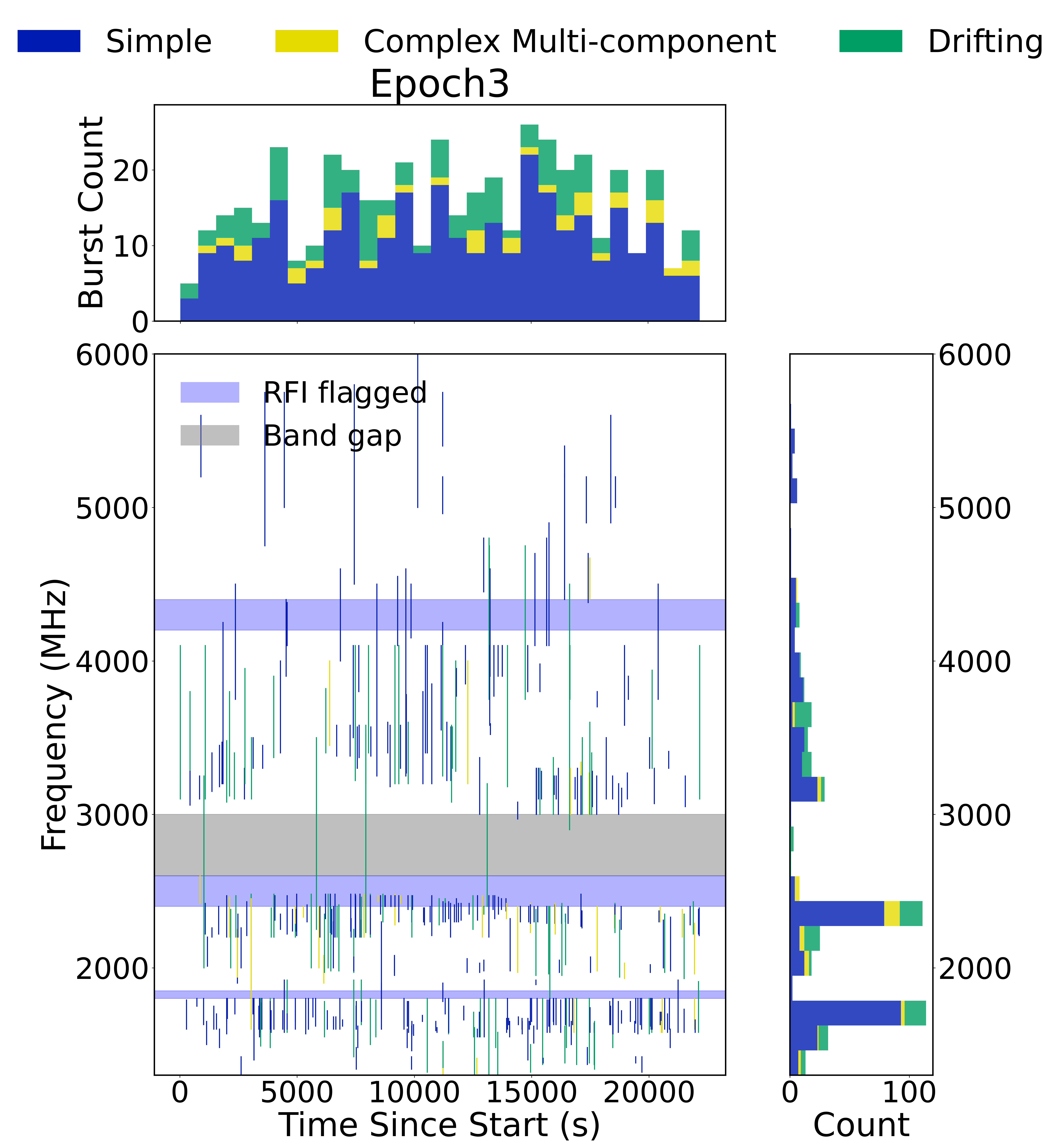}
    \hspace{0.01\linewidth}%
    \includegraphics[width=0.29\linewidth]{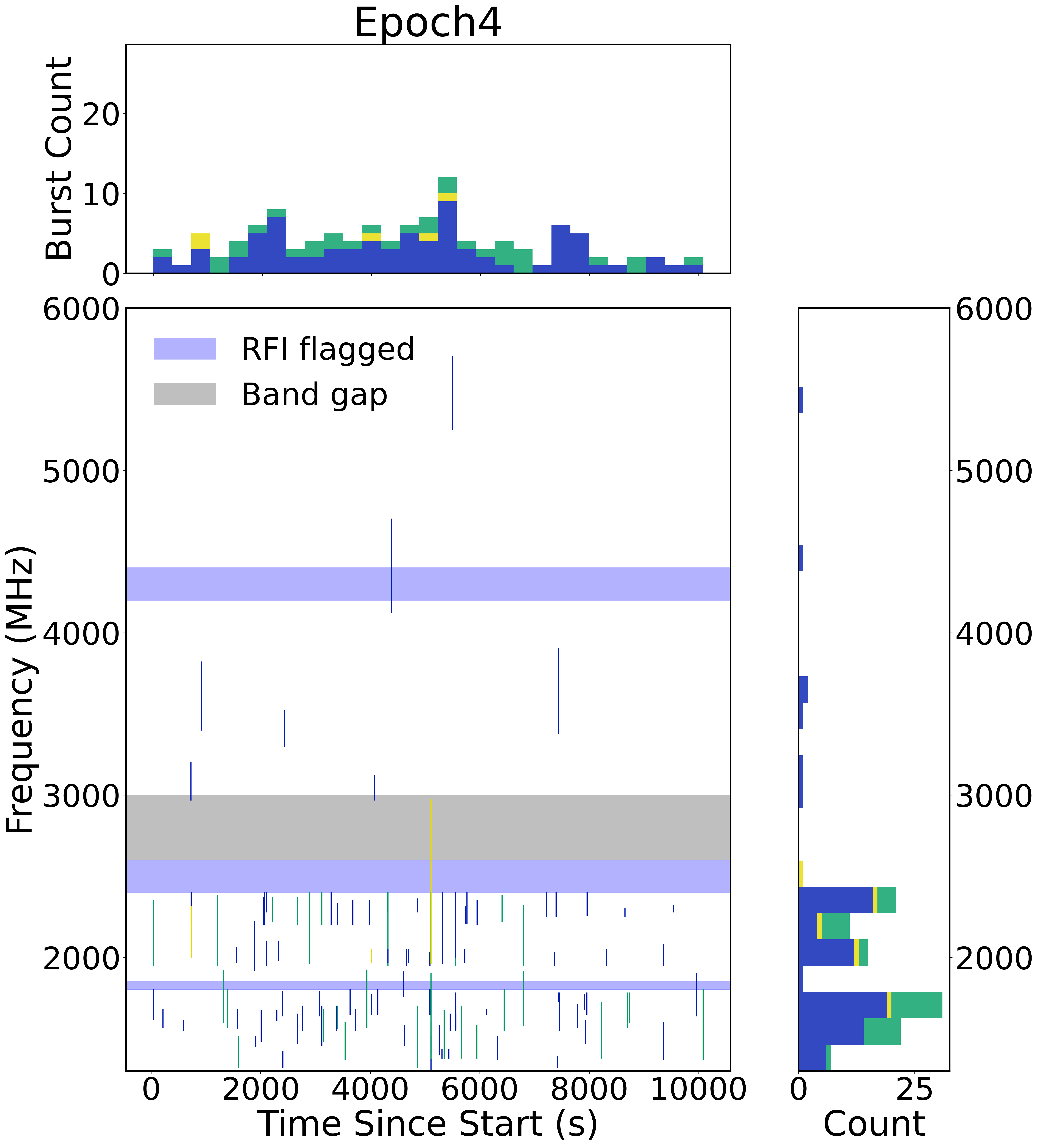}
    \caption{Frequency extent of bursts from \frbA as a function of time for Epochs~2--4. Each vertical line represents an individual burst, color-coded by morphology: simple (blue), complex multi-component (yellow), and drifting (green). The shaded regions indicate flagged frequency ranges due to RFI (light purple) and the instrumental band gap (gray). Side panels show histograms of burst morphologies as a function of time (top) and frequency (right). (We note that an artifical temporal gap is observed in Epoch~2 close to 2000~s due to a technical fault during the observation)}
    \label{fig:three_plots}
\end{figure*}

Figure~\ref{fig:three_plots} summarizes the morphological properties across the last three observing epochs. We note that the first epoch was excluded from this study due to insignificant number of detected bursts. The main panels in this plot show the frequency extent of each burst as a function of time and are color coded by their morphology. The instrumental band gap (shaded gray) and persistently contaminated RFI regions are also marked; these cause apparent cutoffs in burst frequency extents near the edges of the flagged bands. The side and top panels display the spectral and temporal distributions of bursts, grouped by their morphology. Because of the limited number of upward-drifting bursts, we combine upward and downward drifters into a single ``drifting'' category for statistical analysis.

We find no evidence for temporal correlations between burst morphologies and their frequency extents as is visually apparent in Fig.~\ref{fig:three_plots}. Moreover, instrumental effects influence the observed distribution: bursts detected across the band gap and apparent cutoffs near 2.4~GHz arise from RFI contamination. Notably, Epoch~3 shows enhanced burst activity above 4~GHz compared to the other epochs, likely reflecting reduced RFI in this frequency range.

Our morphological classifications are approximately consistent with those in CHIME/FRB Catalog~1 \citep{ziggy}. In particular, CHIME labels type~1 morphology as ``Simple Broadband,'' representing bursts that span the full observing band. In contrast, we find no such broadband bursts in our sample (Fig.~\ref{fig:three_plots}), consistent with previous reports that repeating FRBs tend to emit over narrower fractional bandwidths \citep{ziggy}.

Quantitatively, simple narrowband bursts dominate the sample, comprising $\sim$73\% of all detections. Complex multi-component bursts account for only $\sim$6\%. Downward frequency drifts are observed in $\sim$20\% of the bursts, while rare upward-drifting events ($\sim$1.2\%) are also present. Upward-drifting bursts have previously been reported both in the CHIME/FRB baseband catalog \citep{Faber_chimefrb} and in FAST observations of \frbA \citep{FAST_R147_Morphology}. A frequency-dependent trend is apparent: $\sim$19\% of bursts at lower frequencies (RB1--RB3) exhibit drifts, compared to only $\sim$6\% at higher frequencies, suggesting that drifting structures preferentially occur at lower radio frequencies.

Panel (d) of Fig.~\ref{fig:burst_spectra} illustrates an upward-drifting burst. The full-band view reveals a weak, closely spaced component just below the receiver’s band gap in addition to the upward-drifting sub-structure. Such cases highlight that different drifting morphologies can occur within very short timescales. They also emphasize the importance of wideband observations: had this event been observed with a narrowband receiver, the burst could have been misclassified, leading to an incomplete or biased understanding of its true spectro-temporal structure.

\subsubsection{High-frequency bursts}

In this work, we report detections of emission from \frbA up to 6~GHz. A preliminary report of these high-frequency detections was previously presented in \citet{atel}. As shown in Fig.~\ref{fig:three_plots}, several bursts span the frequency range 5--6~GHz. Notably, the bursts extend up to the edge of our observing band, suggesting that the emission could persist at even higher frequencies.

High-frequency emission has previously been reported from other repeating FRBs. In particular, bursts from \frbB and \frbC have been detected in the 4--8~GHz range \citep{8ghz_frb, R3_Bethapudi}. \citet{8ghz_frb} also identified frequency-drifting structures extending up to 8~GHz. In contrast, the high-frequency bursts detected from \frbA exhibit relatively simple, narrowband profiles. This difference highlights possible variations in propagation effects or intrinsic emission mechanisms among active repeating FRBs.

\subsubsection{Burst widths}

\begin{figure}[h!]
    \centering
    \subfloat{\includegraphics[width=0.49\textwidth]{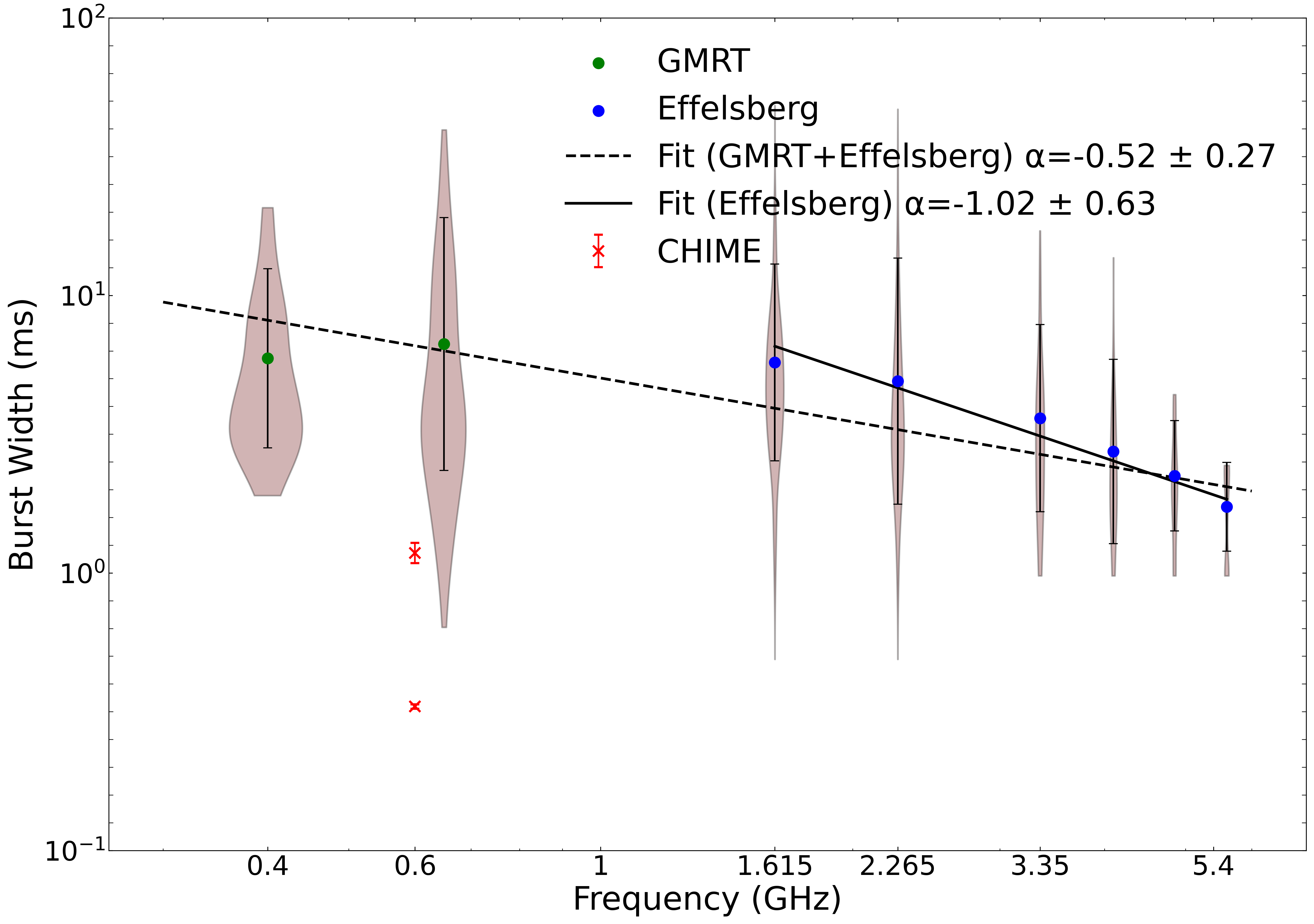}}
    \caption{Frequency evolution of burst widths from GMRT (green circles), Effelsberg (blue circles), and CHIME (red crosses). Distributions at each frequency are shown as violin plots in logarithmic space, with mean values and corresponding uncertainties indicated by black points with error bars. The dashed black line represents the weighted power-law fit across all frequencies (GMRT+Effelsberg), while the solid black line shows the fit using only Effelsberg data. Both fits are performed in log--log space, yielding power-law indices $\alpha$ as indicated in the legend.}

    \label{fig:width_evolution}
\end{figure}

We investigated the frequency dependence of burst duration through the distribution of their temporal widths in each  resampled UBB band, which are depicted with violin plots in Fig.~\ref{fig:width_evolution}. These widths were measured by fitting the coherently dedispersed burst temporal profiles, as described in Section~\ref{sec:burst_properties}. We characterized each distribution by its mean and standard deviation of the bursts and plot these together with violin plots (Fig.~\ref{fig:width_evolution}). As noted in Appendix~\ref{app:SPsearch}, no additional bursts were detected in the highest time resolution search, indicating that the lower tail of the distribution is real and not limited by the time resolution of the data. 

We also compiled the burst widths from a sample of bursts detected by the Giant Meterwave Radio Telescope (GMRT) in observations as reported by \citealp{panda_r147_paper}. We only used widths derived from two observing epochs (out of four), conducted in the 300--500~MHz and 550--750~MHz frequency ranges, as these were the only coherently dedispersed epochs. The GMRT bursts were searched with a lower time resolution limit of $\sim 320~\mu$s, and the search did not extend to the native time resolutions of the different observations. The data were searched up to a maximum temporal width of 32~ms. Hence, a selection bias based on the search methodology cannot be ruled out. Nevertheless, the burst distributions, along with the mean and standard deviation of the widths in each GMRT band, are shown in Fig.~\ref{fig:width_evolution}.

The two blue crosses in Fig.~\ref{fig:width_evolution} indicate the burst widths derived from the two brightest bursts detected by CHIME \citep{r147_chime_paper}. Out of the five CHIME detections, only these two bursts had baseband recordings, which allowed precise width determination using \texttt{fitburst} \citep{fitburst}. Given the low inclination of the source relative to CHIME and its lower sensitivity compared to Effelsberg, CHIME is sensitive only to extremely bright bursts and hence, the two detections are statistically insufficient for a meaningful comparison with widths measured at other frequencies.

A weighted power-law fit of the form \( W \propto \nu^{\alpha} \), where W is the burst width, $\nu$ is the observing frequency and $\alpha$ is the power-law index, was performed in log--log space using only the GMRT and Effelsberg measurements. Each data point was weighted by the standard deviation of the corresponding burst width distribution. The CHIME data were excluded due to the limited sample size and their sensitivity primarily to very bright bursts. The resulting best-fit slope, \(\alpha = -0.52 \pm 0.27\), indicates a moderate inverse scaling of burst width with frequency. However, it remains unclear whether this trend reflects intrinsic properties of the emission source, the emission mechanism, or is significantly influenced by propagation effects such as scattering.  

If we restrict the fit to the Effelsberg sample alone (see Fig.~\ref{fig:width_evolution}), we obtain a steeper slope of $\alpha \sim -1.0 \pm 0.6$, suggesting a stronger frequency dependence of burst width. The apparent flattening seen in the GMRT widths could therefore partly reflect systematic limitations, since their search was restricted to widths up to 32~ms, potentially biasing the recovered distribution. Further broadband observations with uniform temporal resolution and sensitivity will be crucial to establish whether the observed trends are intrinsic or propagation-induced.

Similar frequency-dependent trends have been reported for \frbB \citep{8ghz_frb} and \frbC \citep{R3_Bethapudi}, though these results were obtained from heterogeneous multi-telescope datasets. In contrast, our analysis benefits from a uniform receiver system across all frequencies, thereby reducing systematic uncertainties. We also compared our results with power-law indices derived for other neutron star sources. For instance, broadband observations of the Crab pulsar yielded a power-law index of $\alpha \sim -1.7 \pm 0.1$ \citep{Hankins_2015}, while the broadband radio observations of the magnetar J1622-4950 conducted by \citealp{levin2012} exhibited $\alpha \sim -1.2 \pm 0.3$. These comparisons suggest that coherent emitters exhibit distinct spectral evolution of their temporal widths. Although this does not directly constrain differences in emission mechanisms, it points to the influence of varying emission conditions. Further high-cadence, broadband observations will be essential to clarify the physical processes governing burst width evolution in FRBs.

\subsubsection{Fractional bandwidths}

We define the bandwidth of a burst as the frequency range between its visually identified upper and lower frequency extents of emission. The fractional bandwidth is then given by the ratio of this bandwidth to the burst's center frequency. This provides a useful metric for comparing spectral occupancy across different observing bands.

\begin{figure*}[t]
    \centering
    \includegraphics[width=0.9\linewidth]{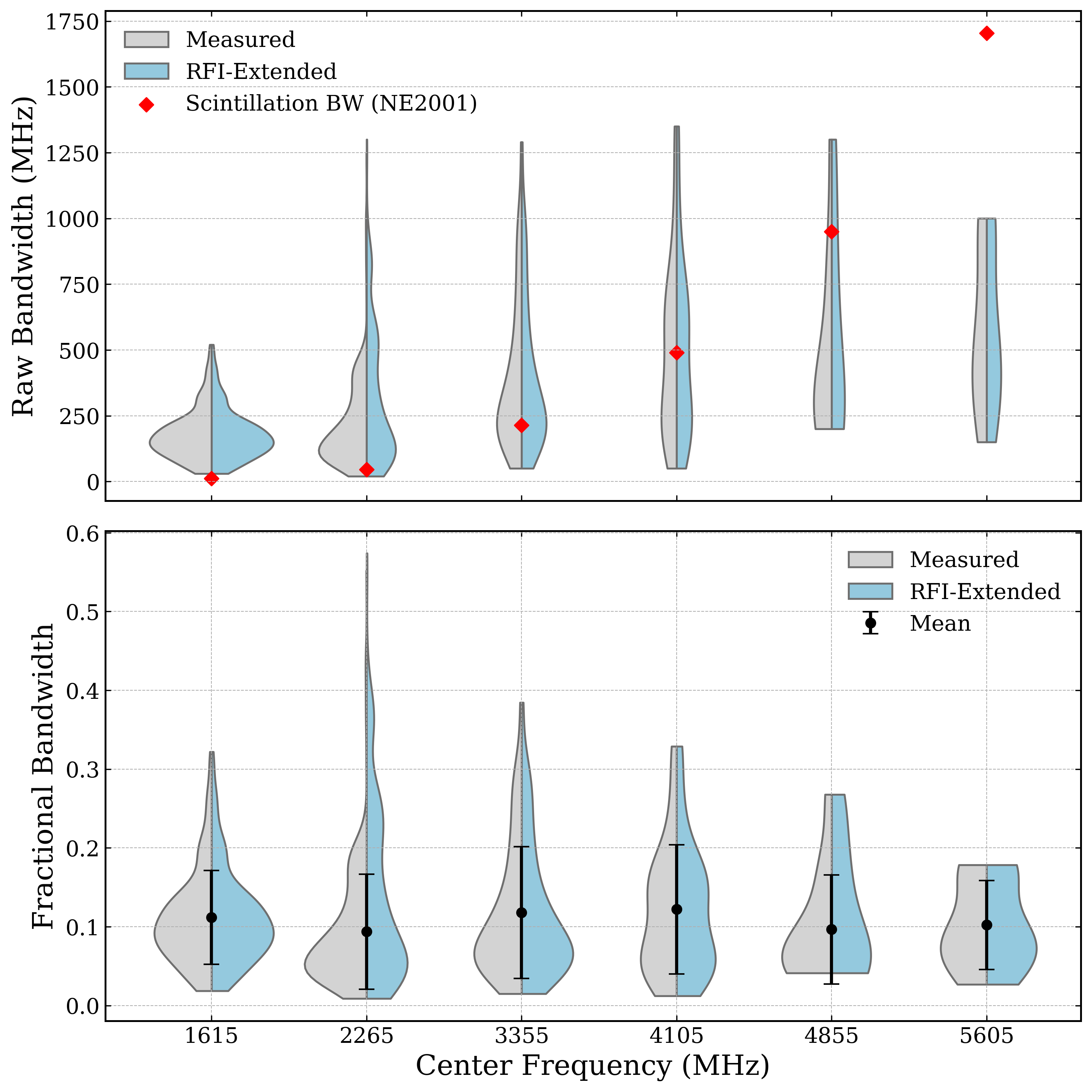}
    \caption{%
        \textbf{Top:} Distribution of raw burst bandwidths in each frequency band shown as violin plots. Gray violins correspond to measured bursts, while blue violins show RFI-extended bursts. The red diamonds mark the scintillation bandwidth predicted by the NE2001 model. 
        \textbf{Bottom:} Fractional bandwidth distributions (burst bandwidth divided by center frequency).
        Gray violins correspond to measured bursts, while blue violins show RFI-extended bursts. 
        Black points with error bars indicate the mean $\pm$ standard deviation for the measured bursts.%
    }
    \label{fig:bw_distribution}
\end{figure*}

The fractional bandwidths of \frbA are shown in Fig.~\ref{fig:bw_distribution}. We present two versions of the distribution in order to account for persistently RFI-contaminated frequency ranges, particularly near the edges of RB2 and RB5. The first, labeled ``Measured,'' is derived directly from the observed dynamic spectra and if the emission ends in an RFI-affected frequency range, the lower edge of this range is assumed. The second, termed ``RFI-extended,  we assume their bandwidth extends to the top edge of the affected regions. In other words, ``Measured" is a lower limit and ``RFI-extended" is the upper limit. In RB2, this shifts the peak in the fractional bandwidth distribution from about 20\% to 30\%, indicating that the initial peak was biased downward by RFI flagging. Similarly, in RB4, the RFI-extended definition produces a smoother distribution compared to the measured one, which drops more sharply due to artificial truncation at the RFI bands.

For comparison, known repeating sources such as \frbC and \frbB typically exhibit fractional bandwidths in the range of 10--30\% \citep{ziggy, Gourdji_2019}. The bursts from \frbA consistently favor  fractional bandwidth occupancy of $\sim$~10\% across a broad frequency range (see Fig.~\ref{fig:bw_distribution}), in agreement with these previous observations of repeaters. This contrasts with apparent non-repeating FRBs, which tend to show broader fractional bandwidths (\citealp{ziggy}). The physical origin of this preferred spectral behavior remains unclear, and future simultaneous broadband observations of both repeating and non-repeating FRBs will be instrumental in probing this empirical trend.

\subsection{Broadband burst rate}
\label{sec:burst_rate}

Broadband observations with the UBB receiver enabled a spectral study of burst rates across a wide frequency range. We define the burst rate as the number of detected bursts per hour in a given frequency range, computed by normalizing the total burst count by the effective source time across all frequency sub-bands. To account for the potential underestimation of the true burst rate, particularly at lower frequencies where RFI significantly reduces search sensitivity, we injected synthetic FRB signals into the filterbank data using the \texttt{FRB Faker} tool\footnote{\url{https://gitlab.com/houben.ljm/frb-faker}}. The bursts were injected with the following parameters: S/N ratios of 7, 10, 30, or 100; widths of 0.1~ms, 1~ms, or 10~ms; and emission bandwidths of the full subband or one quarter of the subband centered at either the top quarter, second quarter, third quarter or bottom quarter. After processing the injected data with \texttt{TransientX}, we found that more than  80\% of the simulated bursts across all the injected burst types were successfully recovered across all UBB bands for ${\rm S/N} \gtrsim 10$.

In order to perform an unbiased comparison across the rebinned bands, we computed the fluence completeness limits in each UBB frequency band for a burst with a typical width of 1~ms, employing the radiometer equation for single pulses (see Appendix~\ref{app:radiometer}). We assumed a detection threshold of ${\rm S/N} \gtrsim 10$ estimated above through burst injection test and the interpolated SEFDs. We limited the impact of persistent interference by identifying the most frequently RFI-flagged frequency channels from a $\sim$30-min \texttt{PSRFITS} file for each UBB band per epoch. A channel was flagged as RFI-affected if it was consistently identified as a RFI-outlier for more than 70\% of the data. Using this list, we redefined the effective bandwidth for each band by excluding the outlier RFI-flagged channels. Finally, we assumed a fiducial burst duration of 1~ms. A script to perform such analysis can be found in a dedicated git repository. \footnote{\url{https://gitlab1.mpifr-bonn.mpg.de/eff-sp-searching/rfiweather}}

\begin{figure*}[h!]
    \centering
    \includegraphics[width=0.99\textwidth]{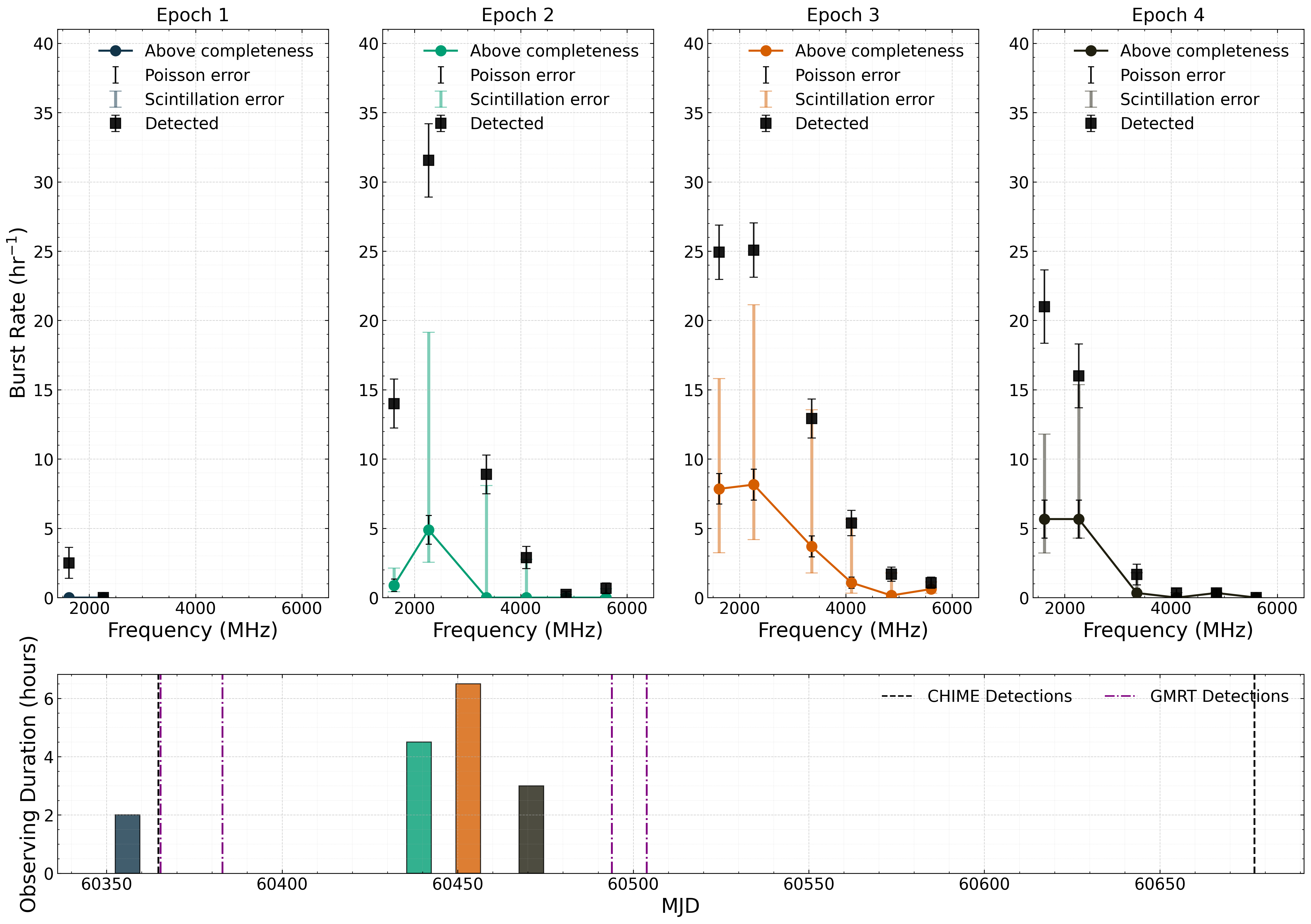}
    \caption{Top panel: The normalized burst count (per hour) as a function of frequency for the five UBB sub-bands over four observing epochs. For each epoch, two rate distributions are plotted: one calculated with all detected bursts and the other restricted to bursts above a common fluence threshold across all sub-bands (see Section~\ref{sec:burst_rate} for details). Error bars represent uncertainties estimated from Poisson statistics, with additional error ranges overplotted to reflect expected variations due to scintillation effects (see text for details). Bottom panel: Total on-source time in each observation as a function of MJD for the four UBB epochs. Vertical dashed lines indicate detection epochs reported by CHIME (black) and uGMRT (blue).}
    \label{fig:burst_rate}
\end{figure*}

In Table~\ref{table:detections} as well as in all other results, we only consider bursts above their respective fluence completeness thresholds. The fluence completeness values for each observing band are reported in Table~\ref{table:rebinned_bands}. For a consistent analysis, we adopted the highest fluence completeness value of $\sim$0.1484~Jy\,ms (corresponding to RB6) as the reference fluence threshold for comparing burst rates across similar energetics over the entire UBB frequency range. In Fig.~\ref{fig:burst_rate}, we show the burst rates above the common fluence completeness threshold as a function of the observing frequency. We also show the burst rate from the detections, without any completeness thresholds to depict the significant variation in the rates after taking into account a common completeness threshold. The error bars assume Poisson statistics. Hereafter, we only consider burst rates above completeness when interpreting the results.

As seen in Fig.~\ref{fig:burst_rate}, a low burst rate was detected in Epoch~1 with no bursts above the completeness threshold. Note that data from the four highest bands were lost from our storage, so no rates are reported. In Epoch~2, the burst rate increased significantly, particularly in the lower-frequency bands (RB1 and RB2). Notably the activity peaked near 2~GHz with burst rate twice as high as $<$2~GHz. Moreover, the 2~GHz peak rate is statistically inconsistent with the neighboring bins assuming Poisson statistics.
Further temporal evolution was seen in Epochs~3 and 4, indicating a shift in the frequency at which peak burst activity occurs. Epoch~3 showed similar rates in RB1 and RB2 but similar rates $>$3~GHz to Epoch 2, while Epoch~4 shows tentative peak in RB1 with significantly fewer bursts detected at high frequencies.

Scintillation, which manifests as a frequency-dependent modulation of source brightness, could also impact the measured burst rate. Scintillation can (de-)magnify the bursts' flux densities, lifting some bursts above detection threshold that might otherwise have been too faint or pushing them below the threshold. The magnification factor depends on the number of frequency scintles ($N_{\rm scint}$) within the emission (or observing) bandwidth and varies on a timescale that depends on the observing frequency and Earth's motion. Statistically the scintillation magnification factor of a burst averaged in frequency follows a $\chi^2$ distribution with $N$$_\mathrm{dof}$ degrees of freedom, where $N_{\rm dof} = N_{\rm scint}$.

The impact of scintillation on the detection of bursts is largest when the scintillation bandwidth (\scintBW ) is comparable to or larger than the emission bandwidth ( \emitBW ) and becomes negligible when a large number of scintles are averaged over. 
For most FRBs scintillation is dominated by the Milky Way foreground, and indeed, a burst from \frbA detected at L-band with the Nan\c{c}ay Radio Telescope shows a marginally resolved scintillation that is consistent with predictions from the NE2001 model \cite[][and reference therein]{r147_chime_paper}. In the Effelsberg sample of bursts, we visually observe spectral structure consistent with scintillation and a detailed characterization is left for future work. 

 The predicted scintillation bandwidth for the line of sight to \frbA, based on the NE2001 model \citep{ne2001}, ranges from 12~MHz at 1.6 GHz (center of RB1) to 1700~MHz (center of RB6), assuming a frequency scaling of $\nu^4$. Comparing these to the observed emission bandwidths (see Fig.~\ref{fig:bw_distribution}), we see that \emitBW /\scintBW\ is $\sim$13 in RB1, $\sim$4 in RB2, and  \emitBW $\sim$ \scintBW\ in RB3 to RB 6. Therefore, scintillation could have an impact on our observed rates, especially at $>$3~GHz. To estimate the impact, we determine the range of magnification factors based on the $\chi^2$ distribution with the appropriate number of degrees of freedom. For example, the magnification from scintllation in RB3 to RB6 follows an exponential distribution ($N_{\rm dof} = N_{\rm scint} =1$), and the 68\% confidence interval corresponds to magnifications between 0.04 and 2. Next we make the assumption that the completeness fluence ($F_{\rm comp}$) is comparable to the mean, intrinsic fluence. In order to alleviate some of the time variability due to scintillation, we combine the burst samples in each UBB band for Epochs 2-4 and re-estimate the rates assuming 0.04 $F_{\rm comp}$ and 2 $F_{\rm comp}$. 
 
 These scintillation (de-)boosted rates are displayed in Fig.~\ref{fig:burst_rate} as the colored errorbars on the fluence complete burst rates, and it is clear that the impact of scintillation can be significant, even in the lower frequency bands. At $\gtrsim$3~GHz the upper limit to the scintillation-boosted rate recovers the rate estimated using all observed bursts in most bands. In addition, the tentative spectral turnover at $\sim$2~GHz could be caused scintillation if the magnification factor boosted the observed bursts' fluences, but it more likely that the turnover is actually more prominent than the fluence completeness rates suggest. We again reiterate that these estimate rely on the highly uncertain assumption that the value of the completeness fluence is similar to the true, intrinsic fluence. Therefore, these results should not be over-interpreted quantitatively but rather are meant to illustrate the potential impact of scintillation on the observed rates. A more detailed study of this is left to future work.

Now, for the purposes of discussion, we assume that the peak in the burst rates at 2~GHz during Epoch~2, as well as the overall temporal evolution of the burst rates, reflects genuine variations in the source spectrum rather than time-variable RFI or scintillation. This behavior resembles the gigahertz-peaked spectra (GPS) commonly observed in some pulsars (\citealp{Kijak_turnover}) and in the magnetar XTE~1810$-$19 \citep{Kijak_turnover2,  Maan_magnetar}. A plausible explanation for such GPS-like behavior is free--free absorption by dense, ionized gas in the FRB environment \citep{Maan_magnetar}. Possible contributors include pulsar wind nebulae, supernova remnants, or HII regions—structures that have also been suggested as origins of the persistent radio sources (PRS) associated with some repeating FRBs \citep{R1_PRS, PRS_190520}. Indeed, \citet{R147_PRS} reported a PRS associated with \frbA, with evidence for spectral steepening in the 1--5~GHz range, while \citet{r147_flaring} observed this source with a peak flux density near 2~GHz. Although the latter does not coincide with the peak burst rate observed by Effelsberg, it highlights the value of coordinated continuum and time-domain observations to probe the PRS--FRB connection.  

At the same time, intrinsic spectral evolution remains a viable alternative. Magnetospheric processes could naturally drive temporal variations in the emission frequency, and evolving magnetospheric conditions may shift the emission to different frequency ranges. Under this scenario, the observed frequency-dependent burst activity would primarily reflect intrinsic emission physics rather than propagation effects.

An extensive follow-up campaign with FAST in the 1.1--1.5~GHz range reported the highest burst rate ever observed for any FRB, reaching $\sim$729 hr$^{-1}$ during one epoch \citep{R147_WaitTime_FAST}. Multiple episodes of heightened activity were recorded, though no clear long-term trend in burst rate evolution was identified. Notably, during the peak of high-frequency activity observed in Epoch~2 with Effelsberg, both the Nançay Radio Telescope and the Allen Telescope Array also reported burst detections above 2~GHz \citep{atel_nancay, atel_allentelescope}. These results indicate that episodes of extreme activity can span a broad frequency range, underscoring the importance of simultaneous multi-band monitoring to capture the full variability of FRB emission.

\subsection{Burst waiting times}
\label{sec:wait_times}

We now investigate the statistics of arrival and waiting times between consecutive bursts in our sample and their possible relation to spectral evolution. For this analysis, we focus on the first three rebinned UBB bands (RB1--RB3), in which more than 100 bursts were detected per band in total in Epochs 2 to 4, improving our statistics. We consider only the arrival times of distinct bursts when analyzing wait time statistics, excluding sub-components within individual bursts. For comparison, the time separations between sub-burst components are shown using distinct histograms as depicted in Fig.~\ref{fig:in_band_waits}.
\subsubsection{Arrival-Time Statistics}
\label{sec:arrival}
If the emission time of a burst is independent of any other emitted burst, the arrival times are expected to follow a Poisson distribution, while bursts whose arrival time is causally related through some process in the emission region may deviate from Poisson.  We quantify the arrival-time statistics through the Weibull probability density distribution with the parameterization given in \citet{oppermann18}:

\begin{equation}
\label{eq:weibull}
    W(\delta | k, r) = k \delta^{-1} \left[ \delta r \Gamma(1+1/k) \right]^k e^{\left[ \delta r \Gamma(1+1/k) \right]^k },
\end{equation}
where $\delta$ are the arrival-time intervals between consecutive bursts, $k$ is the Weibull clustering parameter, $r$ is a constant burst rate, and $\Gamma(x)$ is the gamma function. A Weibull distribution with $k=1$ reduces to a Poisson distribution with a rate parameter $r$. So, our primary test is whether the fitted $k$ values are statistically consistent with $k = 1$. For that, we used a slightly modified version of the  \texttt{frb\_repeation}\footnote{\url{https://github.com/lgspitler/frb_repetition}} code from \citet{oppermann18}.

We fit the burst samples from individual rebinned bands and epochs separately, using only samples with $>30$ bursts.  Therefore, only the burst samples in RB1 and RB2 in Epochs 2, 3, and 4 and RB3 in Epochs 2 and 3 were included. (There was an observing discontinuity in Epoch 2 due to a technical glitch and hence we excluded the bursts before this gap to avoid artificially introducing clustering.) Although \citet{cruces2021} showed that including intervals on short timescales can bias the statistics, we include all bursts in our fits. The resulting fits can be seen in Table~\ref{table:rates}. The burst arrival times in RB1 and RB3 for all epochs considered, as well as RB2 in Epoch 3 are consistent $k=1.0$ at the 1-$\sigma$ level, while in RB2 for Epochs 2 and 4, the consistency is at 2-$\sigma$ and 3-$\sigma$ level, respectively. Therefore, in RB2 there is a greater indication of non-Poisson arrival-time statistics, but as we discuss below, this is the only band where we see short-timescale inter-burst clustering. Therefore, it is likely that these clustered bursts are pulling the fits away from the Poisson expectation.

\subsubsection{Intra-band waiting-time distributions}
\label{subsec:wait_time_observed}

\begin{figure*}[h!]
    \centering
    \includegraphics[width=0.99\textwidth]{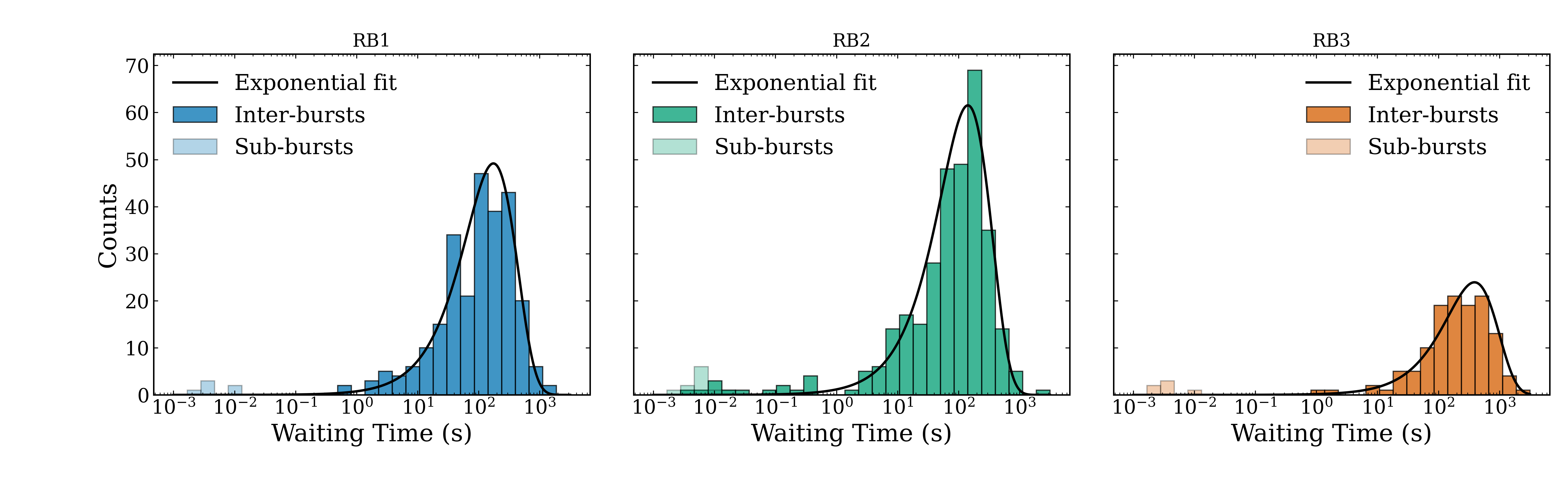}
    \caption{Waiting-time distributions of bursts in the first three rebinned frequency bands (RB1--RB3, left to right). RB1 includes epochs~1--4, while RB2 and RB3 include epochs~2--4. The solid black curve shows an exponential fit to the data using burst rates from Table~\ref{table:rates} for k=1 in Equation~\ref{eq:weibull}. Transparent histograms indicate timescales corresponding to separations between sub-burst components of multi-component bursts.}
    \label{fig:in_band_waits}
\end{figure*}

The waiting time is defined as the time interval between two consecutive bursts within an observation, where each burst occurrence is taken to be the peak time of the highest-amplitude Gaussian component fitted to the burst profile. The waiting-time distributions for each of the three bands with all epochs combined are shown in Fig.~\ref{fig:in_band_waits} and show signs of a bi-modality particularly, in RB2. The longer timescale peak (several 10s of seconds) is likely associated with bursts whose emission times are independent and therefore are expected to follow an exponential distribution. We test this by defining a combined exponential distribution by summing individual exponentials with rates given by the epoch-measured rate in Table~\ref{fig:burst_rate}. These estimated exponential distributions are shown by the solid black curve in Fig.~\ref{fig:in_band_waits}, and it is clear that they describe the data well. To be clear, these were not fits to the histogram, but rather estimated from our rate measurements in Section~\ref{sec:arrival}. Therefore, we associate the longer wait time peak with independently emitted bursts. 

The shorter waiting time peak, with a timescale of $\lesssim 1$~s, indicates temporal clustering of bursts, distinct from the Poissonian process expected for statistically independent events as followed by the long waiting time bursts. Moreover, Fig.~\ref{fig:in_band_waits} illustrates the intervals between sub-bursts within a burst and separate bursts (inter-burst) with different transparencies. For RB1 and RB3, the only short time-scale clustering are from sub-bursts, and are characterized by a timescale of a few milliseconds, i.e., similar to the characteristic duration of a single (sub-)burst. While for RB2 both sub-burst and inter-burst clustering is seen on scales from a few ms to of order 1~s with nearly no gap between the Poisson-distributed events and the small-scale clustering. It is clear that there is a continuum of timescales between these two definitions, suggesting that the definition of burst and sub-burst is somewhat arbitrary. Therefore, the short-duration clustering timescale may be a characteristic of the emission process and similarly fundamental to the emission physics as the $\sim$ms-scale bursts.  

Similar behavior of short-timescale burst clustering is reported for other hyperactive repeaters such as \frbB, \frbD  and FRB~20220912A \citep[e.g.,][]{cruces2021, jahns, R67_wait-time, FRB20220912A_Waits}. For \frbA a peak in the short wait time clustering at $\sim$34~ms was measured in FAST observations at 1--1.5~GHz \citep{R147_WaitTime_FAST}. However, no stable periodicity has been detected in the large FAST burst sample ($>11,000$ bursts from \frbA)\citep{R147_PeriodicitySearch_FAST}. Notably the large sample of bursts shows a bridge between the two peaks with no gaps, similar to what we observed in RB2. Why we detect no inter-burst clustering in RB1, which overlaps with the top half of the FAST bandpass, but clear inter-burst clustering in our RB2, is currently unclear. The wait time distribution of a sample of bursts detected by the uGMRT observations at 300--700~MHz is also bi-modal with a short timescale peak at $\sim$2~ms but that extends out to $\sim$0.5~s \citep{panda_r147_paper}. The authors did not distinguish between sub-burst and inter-burst clustering, so it is possible that the $\sim$2~ms peak is from sub-burst intervals.

If FRB emission is similar to radio-loud Galactic magnetars, short waiting times may reflect multiple bursts within a single rotational phase window \citep{bause_magnetar}, but unlike the arrival times of the bursts from XTE 1810-197, FAST and our RB2 bursts do not show an obvious gap between the intra-rotation bursts and inter-rotation bursts. In fact, the lack of a clear gap strongly suggests that if the bursts originate in the magnetosphere of a rotating neutron star, they must be emitted over a large fraction of the rotational period. Tentative evidence for such behavior was seen in the magnetar SGR1935+2154 (\citealp{sgr1935_waits}), the only known Galactic source of FRB-like radio bursts.

\subsubsection{Spectral-dependent waiting-time distributions}
\label{subsec:band_difference}

To explore whether waiting-time intervals correlate with changes in emission frequency, we define the "band difference`` as the integer separation between rebinned frequency bands in which the centers of consecutive bursts occur. Positive (negative) values indicate upward (downward) shifts in frequency, while zero corresponds to bursts detected in the same band. Instruments with limited frequency coverage cannot detect such inter-band shifts, whereas our broadband observations capture bursts across multiple sub-bands, enabling direct study of frequency evolution with time.

\begin{figure*}[h!]
    \centering
    \includegraphics[width=0.99\textwidth]{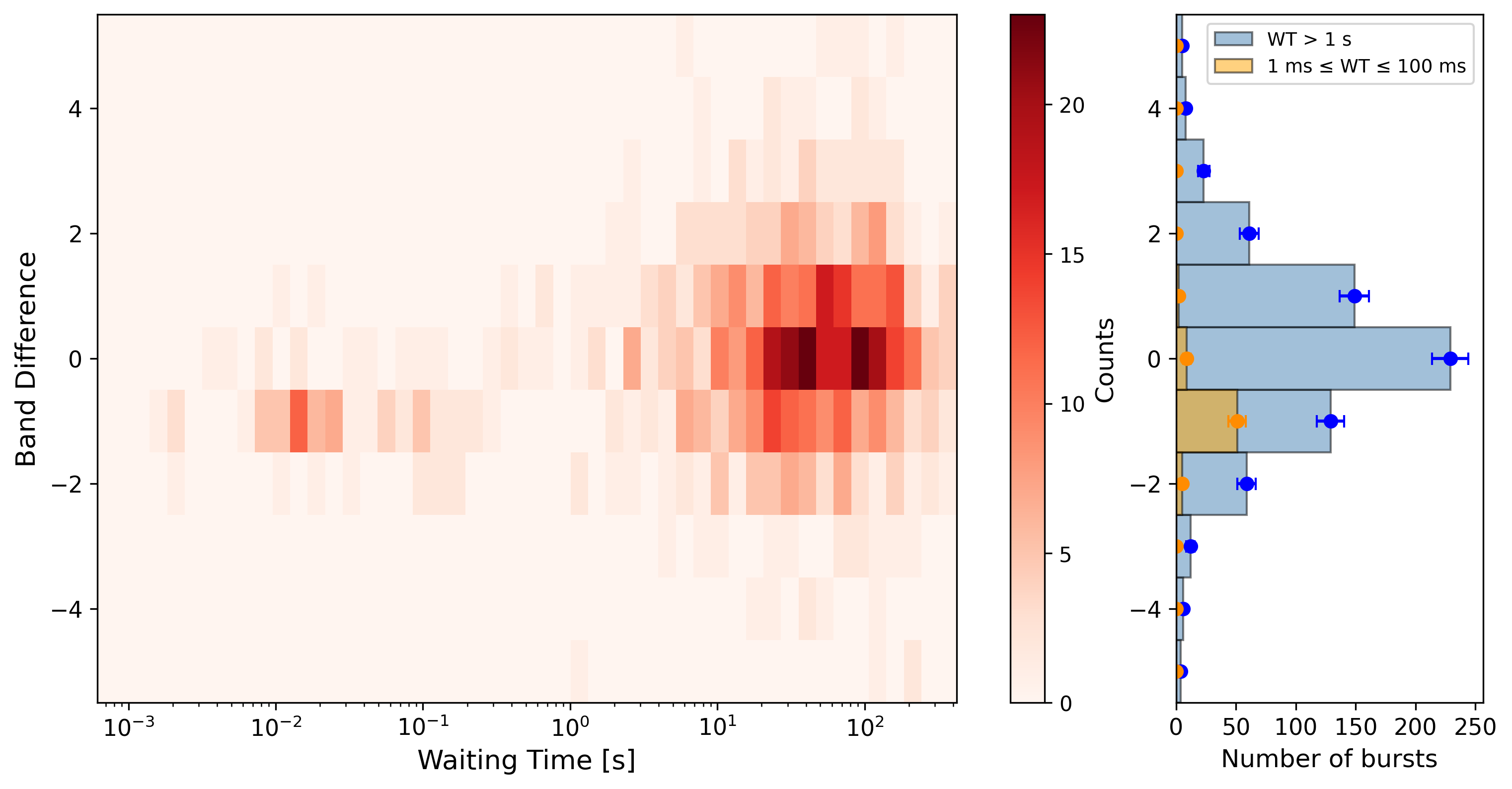}
    \caption{2-D histogram (heatmap) of burst waiting times versus band differences. Waiting times are shown on a logarithmic scale, while band differences indicate jumps between observing bands. Positive values correspond to upward shifts in frequency, while negative values correspond to downward shifts. The long waiting time cluster ($\sim$1--100~s) reflects a Poissonian process with symmetric frequency drifts, while short waiting times ($\lesssim 0.1$~s) show asymmetric band differences dominated by downward jumps. The vertical side panel shows time consolidated waiting-time histogram for bursts shifting towards positive (negative) frequency directions. The blue histogram corresponds to the distribution for the wait times described by a Poissonian process while the orange plot shows statistics in the short wait time cluster of bursts. The Poissonian errors on each waiting-time histogram is depicted by the scatter errorbars.}
    \label{fig:multi_freq_waits}
\end{figure*}

Figure~\ref{fig:multi_freq_waits} presents a two-dimensional distribution of waiting time on the band difference. These band-resolved wait times illustrate the time-scale of frequency jumps between consecutive, independent bursts. For example, a band difference of $\pm$2 or $\pm$5 corresponds to a frequency jump on the order of $\sim$$\pm$1.5~GHz and $\sim$$\pm$4~GHz, respectively. From Fig.~\ref{fig:multi_freq_waits} we see that frequency jumps of $\pm$2 or $\pm$5 bands can occur on time scales of $\sim$second or 10s of seconds, respectively. Therefore, the emission mechanism must be agile enough to switch frequencies separated by $\gtrsim$GHz on time scales of several seconds, and a frequency of $\sim$700\,MHz in a time scale of milliseconds. 

Figure~\ref{fig:multi_freq_waits} also shows two distinct peaks in wait time. The long waiting time cluster ($\sim$1--100~s), which we demonstrated above to be consistent with Poisson statistics for intra-band waiting times, shows a nearly symmetric distribution of upward and downward band jumps, indicating that frequency jumps are uncorrelated on time scales $\gtrsim$1~sec. This is confirmed when the 2D histogram is consolidated along the time axis (see vertical side panel in Fig.~\ref{fig:multi_freq_waits}), which shows that the distribution of band differences for waiting times $>1$~s is symmetric in positive and negative jumps. 

In contrast, bursts with short waiting times ($\lesssim$0.1~s) exhibit a preference for negative band differences, i.e., downward frequency drifts. Importantly, these are not merely sub-burst components but include distinct bursts detected in different frequency bands. This demonstrates that downward frequency evolution is not restricted to intra-burst structure but is also a characteristic of inter-burst clustering. This is further evidence that sub-burst and inter-burst clustering are continuum of the same emission process. 

The preference for downward band differences in clustered bursts suggests evolving plasma conditions in the emission region, such as decreasing particle energies or weakening magnetic fields, as it might occur in expanding magnetized plasma structures or shocks \citep{Wang2024, Vanthieghem2025}. Another plausible scenario involves FRB emission originating from synchrotron maser processes at ultra-relativistic magnetized shocks, where the deceleration of the blast wave leads to a temporally decreasing peak frequency. This naturally results in the observed downward frequency drifts in the FRB emission \citep{metzger_synchrotron_model}.

Alternatively, if the observed FRB emission is significantly affected by plasma lensing near the source, for example, by a supernova nebular shell or a companion's wind, the chromatic magnification of the intrinsic emission is expected. \citet{main_plasma_lensing} demonstrated that extreme plasma lensing from the ionized wind of a companion can induce both upward and downward frequency-drifting structures in the observed pulsar emission, caused by fluctuations in the lensing medium and changes in its geometry relative to the observer. Although such process can alter the source's frequency-time structure and intensity, it remains unclear why the downward band differences are the preferred trend. 

In \frbA, we do observe a small fraction ($\sim$6\%) of bursts exhibiting upward frequency drifts within temporally clustered sequences. However, if plasma lensing were the dominant mechanism, one would expect an approximately equal occurrence of upward and downward drifting bursts. At present, we are not aware of any specific model that explains the upward-drifting bursts observed in FRBs, and therefore we refrain from drawing firm conclusions in this regard.

\section{Summary and Conclusions}
\label{sec:conclusion}

We have presented a comprehensive wide-band study of over 700 bursts from \frbA, spanning from 1.3 to 6~GHz. The key results are summarized below:

\begin{itemize}
    \item Wideband data improve burst detectability and classification; narrowband observations may misclassify drifting bursts and underestimate rates.
    \item No burst was detected extending across the full 1.3-6 GHz band simultaneously. 
    \item Burst morphology shows significant frequency evolution, with complex, frequency-drifting structures more common at lower frequencies. 
    \item The fractional bandwidth remains roughly constant with frequency, while spectral occupancy broadens at higher bands.
    \item The burst durations in time were found to be only weakly dependent on the emission frequency, if at all, being inconsistent with the propagation-induced smearing.
    \item Burst detection rates vary strongly with both frequency and time, peaking in the 1.6--3.5~GHz range. Scintillation is likely contributing to these variations, especially at $>$3~GHz. A tentative spectral turnover is observed around 2~GHz in Epoch~2, but it is unclear whether this is intrinsic to the source, due to the time-variable free-free absorption along the line-of-sight, or scintillation.  
    \item The waiting-time distribution shows bimodality. At longer wait times, the burst arrival times are consistent with a Poisson distribution, which was confirmed with Weibull fits that yield $k \approx 1$, suggesting these events are emitted independently. At shorter wait times, we see a continuum between sub-bursts and inter-burst, likely representing a characteristic timescale of the underlying emission physics that is on the order of 100~ms.
    \item A waiting-time distribution in both time and frequency shows that at shorter timescales, there is a preference for downward drifts, while there is no frequency correlation in arrival times for bursts emitted independently.  
    
\end{itemize}

These results confirm \frbA\ as an active, broadband repeater shaped by both intrinsic and extrinsic processes. Continued wideband, high-time-resolution, and polarimetric observations, alongside theoretical modeling, will be key to uncovering the origin of repeating FRBs.

\section{Acknowledgements}
This work is based on observations with the 100-m telescope of the MPIfR (Max-Planck Institute für Radioastronomie) at Effelsberg. The authors would like to thank Dr. Alex Kraus for scheduling observations with the new UBB receiver. The Effelsberg UBB receiver and EDD backend system are developed and maintained by the electronics division at the MPIfR. PL thanks Suryarao Bethapudi for their initial support in conducting observations and useful discussion during processing the data. LGS is a Lise Meitner Research Group Leader and acknowledges funding from the Max Planck Society. J.B. acknowledge the support by the German Science Foundation (DFG) project BE 7886/2-1. FE and MK acknowledge support from the Deutsche Forschungsgemeinschaft (DFG, grant 447572188).

\section{Data availability}
The data underlying this article will be shared on reasonable request to the corresponding authors.

\bibliographystyle{aa} 
\bibliography{refs}
\clearpage

\appendix
\setcounter{table}{0}
\setcounter{figure}{0}
\renewcommand{\thetable}{\arabic{table}}
\renewcommand{\thefigure}{\arabic{figure}}
\section*{Appendix}
\renewcommand{\thesubsection}{\Alph{subsection}} 
\FloatBarrier

\subsection{Single pulse search details}
\label{app:SPsearch}

For the first two epochs, the data were downsampled in time by a factor of 8, and for the last two epochs by a factor of 4, for the width search from 1~ms--40~ms in order to match the different native resolutions of the data. To mitigate narrowband RFI, the data were further downsampled in frequency by a factor of 4. We also carried out searches up to the native time resolutions to test for the presence of very narrow bursts, but no unique detections were found in this regime.

RFI mitigation in \texttt{TransientX} included the kadane filtering algorithm to suppress narrowband interference, complemented by manual masks to remove persistently contaminated frequency ranges. For the first three sub-bands, the \texttt{zdot} option was enabled to apply a zero-DM filter, removing broadband signals centered at zero dispersion measure. Additionally, zapped RFI channels were replaced with the mean of each corresponding 1~s time segment to preserve the statistical properties of the data.

Dedispersion was carried out over a DM range of 500--550~pc\,cm$^{-3}$, with step sizes computed using \texttt{DDplan.py}\footnote{\url{https://github.com/scottransom/presto/blob/master/bin/DDplan.py}}. 

Finally, the DBSCAN clustering algorithm was used with a radius of 1~ms to merge duplicate detections across widths and adjacent trial DMs. Each resulting cluster was refined with \texttt{replot\_fil}, which performs a high-resolution reprocessing to reduce false positives before visual inspection.

\subsection{Methodology of burst properties extraction}
\label{app:burst_appendix}

The burst times-of-arrival (TOAs) were extracted from \texttt{TransientX} candidates identified in each sub-band. To ensure consistency across frequency bands, all data were dedispersed to a fixed DM of 527.979~pc~cm$^{-3}$, obtained from a structure-optimized fit to a bright burst \citep{eppel_r147}. Single-pulse archives of 1-second duration were generated using \texttt{DSPSR}\footnote{\url{https://dspsr.sourceforge.net/}} for all sub-bands.

RFI was mitigated using: (1) manual masks derived from folded 1~s archives, (2) a zero-DM filter for broadband RFI suppression (up to Band~3), and (3) the \texttt{CLFD} algorithm \citep{CLFD}. Flux calibration used on/off scans of 3C48 with a 0.5~s diode switching period, processed using \texttt{PSRCHIVE} tools \texttt{fluxcal} and \texttt{pac}\footnote{\url{https://psrchive.sourceforge.net/}} using the 3C48 calibrator model from \citealp{fluxcal_citation}. 

Calibrated archives were loaded into 2D arrays via the \texttt{PSRCHIVE Python Interface}\footnote{\url{https://psrchive.sourceforge.net/manuals/python/}}. Dynamic spectra and temporal profiles were reconstructed as shown in Fig.~\ref{fig:burst_spectra}. Due to significant RFI in lower-frequency bands, bandwidths were visually estimated instead of fit-based measurements.

Burst arrival times were referenced to the center frequency of the highest band in which the burst was detected (rather than the highest channel overall, as in the initial detection). This referencing allowed us to properly handle cross-band burst detections without duplication.

Temporal structure was characterized using a multi-Gaussian fitting approach. We first smoothed the time series to identify statistically significant peaks (above a fixed S/N threshold), then fitted Gaussian components using \texttt{scipy.optimize.curve\_fit} \citep{2020SciPy-NMeth}. Gaussian components were iteratively added until residuals converged below a reduced-$\chi^2$ threshold of 0.01.

Two components were considered sub-structures of the same burst if their FWHMs overlapped; otherwise, they were treated as distinct bursts. From the fits, we extracted FWHMs, TOAs (relative to observation start), peak flux densities (Jy), and fluences (Jy--ms). A catalog of derived properties for a representative burst sample is provided in Table~\ref{tab:burst_properties}.

\subsection{Radiometer equation and fluence estimation}
\label{app:radiometer}

The radiometer equation for single pulses is given by
\begin{equation}
    S = \frac{S/N \cdot \mathrm{SEFD}}{\sqrt{n_{p} \cdot \Delta\nu \cdot \Delta W}},
\end{equation}
where $S$ is the signal strength (Jy), $S/N$ is the signal-to-noise ratio, $\mathrm{SEFD}$ is the system equivalent flux density (Jy), $n_{p} = 2$ is the number of polarizations summed , $\Delta\nu$ is the bandwidth (Hz), and $\Delta W$ is the effective pulse width or time resolution (s).

Following this equation, the fluence $\mathscr{F}$ (Jy\,ms) of a single pulse can be estimated as:
\begin{equation}
    \mathscr{F} = \frac{S/N \cdot \mathrm{SEFD} \cdot \sqrt{\Delta W}}{\sqrt{n_{p} \cdot \Delta\nu}}.
\end{equation}

\onecolumn
\subsection{Representative burst properties}
\label{app:burst_table}

\begin{table}[H]
\centering
\begin{adjustbox}{max width=\textwidth}
\begin{tabular}{cccccc}
\toprule
\textbf{TOA (MJD)} & \boldmath$\Delta t$ \textbf{(ms)} & \boldmath$\Delta\nu$ \textbf{(MHz)} & \boldmath$\nu_c$ \textbf{(MHz)} \\
\midrule
60439.1704929 & 3.42 & 70  & 2015 \\
60439.1713895 & 1.46 & 80  & 2340 \\
60439.2091054 & 6.35 & 240 & 1800 \\
60439.2097786 & 9.28 & 220 & 1810 \\
60439.2179350 & 6.84 & 220 & 1810 \\
60439.2179350 & 9.77 & 220 & 1810 \\
60439.2416931 & 2.44 & 130 & 2315 \\
60439.2438170 & 2.44 & 80  & 2240 \\
60439.3655775 & 2.93 & 50  & 1775 \\
60439.3662699 & 8.79 & 170 & 1715 \\
\bottomrule
\end{tabular}
\end{adjustbox}
\caption{Representative sample of burst temporal and spectral properties from \frbA. 
Times of arrival (TOAs) are barycentered and dedispersed to the topmost observing frequency. 
Burst widths ($\Delta t$) are derived from Gaussian fits, while bandwidths ($\Delta\nu$) and central frequencies ($\nu_c$) are estimated through visual inspection. The burst flux densities and fluence will be reported in a paper currently in preparation.}
\label{tab:burst_properties}
\end{table}

\subsection{Details on the receiver specifications}

\begin{table*}[h!]
\centering
\begin{tabular}{lccccc}
\toprule
\textbf{Band} & \textbf{Frequency Coverage} & \textbf{Bandwidth} & \textbf{Frequency Channels} & \textbf{SEFD} \\
 & (MHz) & (MHz) &  & (Jy) \\
\midrule
Band 1 & 1290--1939 & 650 & 1280 & 12.2 \\
Band 2 & 1940--2589 & 650 & 1280 & 11.2 \\
Band 3 & 2976--4101 & 1125 & 2400 & 11.5 \\
Band 4 & 4101--5226 & 1125 & 2400 & 13.8 \\
Band 5 & 5226--5976 & 750 & 1600 & 17.5 \\
\bottomrule
\end{tabular}

\caption{Frequency coverage, channelization, and sensitivity (SEFD) specifications for each sub-band recorded using the Effelsberg UBB receiver and EDD backend. The System Equivalent Flux Densities (SEFDs) were estimated using observations of the standard calibrator 3C48.}
\label{table:rec_details}
\end{table*}
\FloatBarrier

\subsection{Burst widths across different frequencies}

\begin{center}
\renewcommand{\arraystretch}{1.4}
\begin{tabular}{llcc}
\toprule
\textbf{Telescope} & \textbf{Band / Burst} & \textbf{Center Frequency (MHz)} & \textbf{Width (ms)}  \\
\midrule
CHIME & Burst1 & 600 & $1.18 \pm 0.10$ \\
CHIME & Burst2 & 600 & $0.33 \pm 0.006$  \\
\midrule
GMRT & Band 1 & 400 & $3.9 \pm 0.3$  \\
GMRT & Band 2 & 650 & $4.1 \pm 0.2$  \\
Effelsberg & RB1 & 1615 & $4.7 \pm 0.06$  \\
Effelsberg & RB2 & 2265 & $3.7 \pm 0.07$  \\
Effelsberg & RB3 & 3350 & $2.6 \pm 0.2$  \\
Effelsberg & RB4 & 4100 & $2.7 \pm 2.09$  \\
Effelsberg & RB5 & 4850 & $2.2 \pm 1.02$  \\
Effelsberg & RB6 & 5600 & $1.7 \pm 0.6$ \\
\bottomrule
\end{tabular}
\captionof{table}{Summary of burst widths and properties across CHIME, GMRT, and Effelsberg frequency bands. The first two rows present the CHIME-detected bursts with their measured burst widths, based on fits to baseband data using \texttt{fitburst}. The remaining rows list representative mean burst widths across GMRT and Effelsberg sub-bands. The CHIME burst values are adapted from \citet{r147_chime_paper}.}
\label{table:combined_burst_widths}
\end{center}
\FloatBarrier

\subsection{Burst rates and temporal clustering}
\begin{table}[H]
\centering
\renewcommand{\arraystretch}{1.5}  
\begin{adjustbox}{max width=0.85\textwidth}
\begin{tabular}{lcccccc}
\toprule
\multirow{3}{*}{\textbf{Band}} 
 & \multicolumn{3}{c}{\textbf{Burst Rate}} 
 & \multicolumn{3}{c}{\textbf{Clustering Parameter $k$}} \\
\cmidrule(lr){2-4} \cmidrule(lr){5-7}
 & \textbf{Epoch 2} & \textbf{Epoch 3} & \textbf{Epoch 4} 
 & \textbf{Epoch 2} & \textbf{Epoch 3} & \textbf{Epoch 4} \\
 & \multicolumn{3}{c}{($\mathrm{hr}^{-1}$)} & \multicolumn{3}{c}{} \\
\midrule
RB1 & $13.4^{+2.4}_{-2.0}$ & $22.3^{+2.7}_{-2.5}$ & $19.7^{+3.1}_{-2.4}$ & $0.9^{+0.1}_{-0.1}$ & $0.9^{+0.1}_{-0.1}$ & $1.0^{+0.1}_{-0.1}$ \\
RB2 & $28.6^{+3.0}_{-3.5}$ & $23.7^{+2.5}_{-2.5}$ & $14.4^{+3.7}_{-3.0}$ & $0.8^{+0.1}_{-0.1}$ & $0.9^{+0.1}_{-0.1}$ & $0.7^{+0.1}_{-0.1}$ \\
RB3 & $8.2^{+1.7}_{-1.4}$ & $12.6^{+1.1}_{-1.2}$ & -- & $1.0^{+0.1}_{-0.1}$ & $1.0^{+0.1}_{-0.1}$ & -- \\
\bottomrule
\end{tabular}
\end{adjustbox}
\caption{Burst rates and clustering parameters derived from Weibull distribution fits in the first three rebinned UBB frequency bands across three Effelsberg observing epochs. The burst rates are given in units of hr$^{-1}$ and the clustering parameter $k$ is dimensionless. The burst count in RB3 during Epoch~4 was insufficient, so the corresponding parameters were not measured. These rates were used to model waiting times with an exponential distribution in Fig.~\ref{fig:in_band_waits}.}
\label{table:rates}
\end{table}
\FloatBarrier

\end{document}